\documentclass{article}%
\usepackage{amsfonts}
\usepackage{amsmath}
\usepackage{amssymb}
\usepackage{graphicx}
\usepackage{fancybox}
\usepackage{geometry}%
\newtheorem{theorem}{Theorem}

\newtheorem{corollary}[theorem]{Corollary}

\newtheorem{definition}[theorem]{Definition}

\newtheorem{proposition}[theorem]{Proposition}
\newtheorem{remark}[theorem]{Remark}

\newenvironment{proof}[1][Proof]{\noindent\textbf{#1.} }{\ \rule{0.5em}{0.5em}}
\newlength{\querylen}
\begin{document}

\title{Quasi self-dual exponential L{\'e}vy processes \thanks{The authors
are grateful to Ilya Molchanov and Michael Tehranchi for helpful
discussions and hints. Michael Schmutz was supported by Swiss
National Fund Project Nr. 200021-126503. Thorsten Rheinl\"ander
would like to thank the FIM at ETH Zurich at which part of this
work was carried out during a sabbatical stay. Furthermore,
financial support by EPRSC is gratefully acknowledged.}}
\date{}
\author{Thorsten Rheinl\"ander\thanks{Department of Statistics, London
School of Economics, Houghton Street, London, WC2A 2AE, United
Kingdom. (\texttt{t.rheinlander@lse.ac.uk}). }\ \, and Michael
Schmutz\thanks{Department of Mathematical Statistics and Actuarial
Science, University of Bern, Sidlerstrasse 5, 3012 Bern,
Switzerland (\texttt{michael.schmutz@stat.unibe.ch}). }
}\maketitle

\begin{abstract}
  The important application of semi-static hedging in financial
  markets naturally leads to the notion of quasi self-dual
  processes. The focus of our study is to give new characterizations
  of quasi self-duality for exponential L{\'{e}}vy processes such
  that the resulting market does not admit arbitrage opportunities.
  We derive a set of equivalent conditions for the stochastic
  logarithm of quasi self-dual martingale models and derive a
  further characterization of these models not depending on the
  L\'{e}vy-Khintchine parametrization. Since for non-vanishing order
  parameter two martingale properties have to be satisfied
  simultaneously, there is a non-trivial relation between the order
  and shift parameter representing carrying costs in financial
  applications. This leads to an equation containing an integral
  term which has to be inverted in applications. We first discuss
  several important properties of this equation and, for some
  well-known models, we derive a family of closed-form inversion
  formulae leading to parameterizations of sets of possible
  combinations in the corresponding parameter spaces of well-known
  L\'{e}vy driven models.

   \medskip



\end{abstract}

\pagestyle{myheadings} \thispagestyle{plain} \markboth{THORSTEN
RHEINL{\"A}NDER AND MICHAEL SCHMUTZ}{Quasi self-dual exponential
L{\'e}vy processes}

\section{Introduction}

The duality principle in option pricing relates different
financial products by a certain change of measure. It allows to
transform complicated financial derivatives into simpler ones in a
suitable dual market. For a comprehensive treatment,
see~\cite{eber:pap:shir08,eber:pap:shir08b} and the literature
cited therein.

Sometimes it is even possible to semi-statically hedge
path-dependent barrier options with European ones. These are
options which only depend on the asset price at maturity. The
possibility of this hedge, however, requires a certain symmetry
property of the asset price which has to remain invariant under
the duality transformation. Non-vanishing carrying costs like
interest rates, dividends etc.\ are handled in the previous
literature by quasi self-dual processes which remain invariant
under duality after a power transform, see~\cite{car:cho97} and
more recently e.g.~\cite{CL,MoS}. Furthermore, this allows more
modelling flexibility since quasi self-duality is a less
restrictive requirement on the price process compared to classical
self-duality. For references to the large literature of the
special case of self-duality, often called put-call symmetry,
see~\cite{CL,eln:jem99,faj:mor10,T}.

Moreover, quasi-self duality in one period shows up in the study
of probabilistic representations of the Riemann zeta function, see
e.g.\ equation (11.4) in \cite{MY}.

The focus of our study is to give new characterizations of quasi
self-duality for exponential L{\'{e}}vy processes such that the
corresponding market does not admit arbitrage opportunities.
Indeed, it is discussed in~\cite{MoS} that ensuring quasi
self-duality and the absence of arbitrage simultaneously is a
delicate task in many L\'{e}vy settings. We derive a set of
equivalent conditions for the stochastic logarithm of quasi
self-dual martingale models and derive a further characterization
of these models not depending on the L\'{e}vy-Khintchine
parametrization. Furthermore, we complement the characterization
of quasi self-dual martingale models obtained in~\cite{MoS} for
our slightly different definition, where the order parameter of
quasi self-duality is allowed to vanish. Since for non-vanishing
order parameter two martingale properties have to be satisfied
simultaneously there is a non-trivial relation between the order
and the shift parameter representing carrying costs in financial
applications. This leads to an equation containing an integral
term which has to be inverted in applications. We first discuss
several important properties of this equation and for some
well-known models, we derive a family of closed-form inversion
formulae leading to parameterizations of sets of possible
parameter combinations in the corresponding parameter spaces of
well-known L\'{e}vy driven models. Furthermore, we discuss an
example where we do not end up with a unique inversion formula for
possible carrying costs after fixing all parameters up to the
shift (carrying costs) and the order parameter.

Since hedging portfolios in applications, i.e.\ the construction of
semi-static hedging strategies, substantially depend on the order
parameter, this study leads to new explicit semi-static hedging
portfolios in markets without arbitrage. In real market applications
often the assumption that the asset price follows certain
exponential L\'{e}vy processes will typically not be completely
fulfilled and the possibility of jumps will lead to certain hedging
errors. However, several comparative studies, see
e.g.~\cite{eng:fen:nal:schw06,nal:pou06}, have confirmed a
relatively good performance of \textquotedblleft symmetry
based\textquotedblright\ semi-static hedges, even if the assumptions
behind the semi-static hedges are not exactly satisfied.

\section{Definitions and applications}

We work on a filtered probability space $\left(
\Omega,\mathcal{F},\mathbb{F},P\right)  $ where the filtration
satisfies the usual conditions with $\mathcal{F}_{0}$ being
trivial apart from $P$-null sets, and fix a finite but arbitrary
time horizon $T>0$. All stochastic processes are RCLL and defined
on $\left[ 0,T\right]  $ unless otherwise stated. We understand
positive and negative in the strict sense. As far as the
definitions are concerned we follow~\cite{RhS}.

\begin{definition}
\label{def:r-sym} Let $M$ be an adapted process. $M$ is
\textbf{conditionally symmetric} if for any stopping time
$\tau\in\left[  0,T\right]  $ and any non-negative Borel function
$f$
\begin{equation}
E\left[  \left.  f\left(  M_{T}-M_{\tau}\right)  \right\vert \mathcal{F}%
_{\tau}\right]  =E\left[  \left.  f\left(  M_{\tau}-M_{T}\right)
\right\vert
\mathcal{F}_{\tau}\right]  . \label{symmetry condition}%
\end{equation}

\end{definition}

Here it is permissible that both sides of the equation are
infinite. If $M$ is an integrable conditionally symmetric process,
then Condition~(\ref{symmetry condition}) implies that $M$ is a
martingale by choosing $f(x)=x$ ($=x^{+}-x^{-}$).

\begin{definition}
\label{def:qsd} An adapted positive process $S$ is \textbf{quasi
self-dual }of
order $\alpha\in%
\mathbb{R}
$ if for any stopping time $\tau\leq T$ and any non-negative Borel
function
$f$ it holds that%
\begin{equation}
\label{eq:def-quasi-equation}E_{P}\left[  \left.  f\left(  \frac{S_{T}%
}{S_{\tau}}\right)  \right\vert \mathcal{F}_{\tau}\right]
=E_{P}\left[ \left.  \left(  \frac{S_{T}}{S_{\tau}}\right)
^{\alpha}f\left(  \frac {S_{\tau}}{S_{T}}\right)  \right\vert
\mathcal{F}_{\tau}\right]  .
\end{equation}
In particular, for all $\tau\leq T$
\[
E_{P}\left[  \left.  \left(  \frac{S_{T}}{S_{\tau}}\right)
^{\alpha }\right\vert \mathcal{F}_{\tau}\right]  =1.
\]

\end{definition}

Note that even for $\alpha=1$ these definitions differ slightly
from the one used in~\cite{T} who uses bounded measurable $f$
instead. Moreover, in view of applications we use bounded stopping
times instead of deterministic times. However, all corresponding
results in~\cite{T} applied in this paper can be easily adapted to
our corresponding setting as is straightforward to check.

In the case when $S$ is a martingale, we can define a probability
measure $Q$,
the so-called \emph{dual measure}, via%
\begin{equation}
\frac{dQ}{dP}=\frac{S_{T}}{S_{0}}. \label{measure Q}%
\end{equation}
Similarly, if $E\left[  \sqrt{S_{T}}\right]  <\infty$, or $E\left[  S_{T}%
^{w}\right]  <\infty$ for a $w\in\left[  0,1\right]  $,
respectively, we define probability measures $H$, sometimes called
`half measure', respectively
$P^{w}$, via%
\begin{equation}
\frac{dH}{dP}=\frac{\sqrt{S_{T}}}{E\left[  \sqrt{S_{T}}\right]
}\,,\quad \frac{dP^{w}}{dP}=\frac{S_{T}^{w}}{E\left[
S_{T}^{w}\right]  }\,.
\label{measure H}%
\end{equation}

Note that the integrability of $S_{T}=S_{0}\exp(X_{T})$ under $P$
implies the existence of the moment generating function of $X_{T}$
under $H$ for an open interval including the origin, i.e. $X_{T}$
has all moments under $H$.

By Bayes' formula, the self-duality
condition~(\ref{eq:def-quasi-equation}) for $\alpha=1$ can be
expressed for a martingale $S$ in terms of the dual
measure $Q$ defined in~(\ref{measure Q}) as%
\begin{equation}
E_{P}\left[  \left.  f\left(  \frac{S_{T}}{S_{\tau}}\right)
\right\vert
\mathcal{F}_{\tau}\right]  =E_{Q}\left[  \left.  f\left(  \frac{S_{\tau}%
}{S_{T}}\right)  \right\vert \mathcal{F}_{\tau}\right]  .
\label{self-duality numeraire}%
\end{equation}

\medskip

For the measure $H$, i.e.~$w=\frac{1}{2}$, the following
proposition has been stated in slightly different settings
in~\cite{CL,MoS} and~\cite{T}, and also for $w=1$, i.e.\ for $Q$.
Similar unconditional multivariate results are given
in~\cite{mol:sch11}.

\begin{proposition}
\label{symmetry under H-gen} Let $S=\exp\left(  X\right)  $ be a
martingale. Then $S$ is self-dual if and only if for any stopping
time $\tau\in[0,T]$ and any non-negative Borel function $f$
\begin{equation}
E_{P^{w}}\left[  \left.  f\left(  X_{T}-X_{\tau}\right)
\right\vert
\mathcal{F}_{\tau}\right]  =E_{P^{1-w}}\left[  \left.  f\left(  X_{\tau}%
-X_{T}\right)  \right\vert \mathcal{F}_{\tau}\right]
\label{eq:symmetry-under-H-gen}%
\end{equation}
holds for at least one (and then necessarily for all) $w\in[0,1]$.
\end{proposition}

\begin{proof}
See~\cite{RhS}, Proposition 4; note that this proof does not need
that $S$ is continuous.
\end{proof}

\medskip

For the half measure we immediately obtain the known special case,
which was formulated in slightly different settings
in~\cite{CL,MoS,RhS,T}.

\begin{corollary}
\label{symmetry under H} Let $S=\exp\left(  X\right)  $ be a
martingale. Then $S$ is self-dual if and only if $X$ is
conditionally symmetric with respect to $H$.
\end{corollary}

Note that this corollary shows that the self-duality definition
employed here ($\alpha=1$) is equivalent to the assumption that
put-call symmetry at stopping times, as defined in~\cite{CL},
holds for all stopping times $\tau\in\lbrack0,T]$.

The following result is sometimes used as definition or as
starting point in slightly different settings, see~\cite{CL}, who
treat the case of vanishing parameter in concrete models
separately, and~\cite{MoS}. The advantage of
Definition~\ref{def:qsd} is that we do not need to exclude
$\alpha=0$ at the starting point or in the definition. Note
furthermore, that our terminology also minimally differs from the
one in~\cite{MoS} where the part without carrying costs is called
to be self-dual and not the asset price process itself.

\begin{proposition}
\label{characterization of quasi self-duality} $S$ is quasi
self-dual of order $\alpha\neq0$\textbf{\ }if and only if
$S^{\alpha}$ is self-dual.
\end{proposition}

\begin{proof}
This follows by considering for each $f$ the functions $g$ defined
by $g\left(  x\right)  =f\left(  x^{\alpha}\right)  $,
respectively $h$ given by $h\left(  x\right)  =f\left(
x^{1/\alpha}\right)  $, $x>0$.
\end{proof}

\medskip

For $S^{\alpha}$ being a martingale for non-vanishing $\alpha$ and
such that the discounted asset price process is also a martingale, a
machinery of hedging strategies has been derived
in~\cite[Section~6.2]{CL} and is concretely discussed in the
geometric Brownian motion case, there also for vanishing $\alpha$ as
in~\cite{car:cho97}. Certain extensions to geometric Brownian motion
including jumps to zero and to one-dimensional diffusions are
derived in~\cite[Section~7]{CL}, while the structural results
in~\cite{RhS} show that a reasonable calibration of continuous asset
price models leading at the same time to quasi self-duality can
usually not be expected due to symmetry reasons.

For illustration we repeat here one particular hedging strategy,
other ones can be derived from~\cite{CL} in an analogous way.
Consider
\[
X=f(S_{T})1\hspace*{-0.55ex}\mathrm{I}_{\big\{\exists t\in\lbrack
0,T],\;S_{t}\,\leq\,H\big\}}\,,
\]
along with
\begin{equation}
  g(S_T)=f(S_{T})1\hspace*{-0.55ex}\mathrm{I}_{\big\{S_{T}\leq H\big\}}+\Big(\frac
{S_{T}}{H}\Big)^{\alpha}f\Big(\frac{H^{2}}{S_{T}}\Big)1\hspace*{-0.55ex}%
\mathrm{I}_{\big\{S_{T}<H\big\}}\,,\label{eq:hedge-t}%
\end{equation}
such that $f$ and $g$ are non-negative, integrable payoff functions
($S_{0}>H$, $H>0$). Furthermore, assume that $S$ is positive quasi
self-dual of non-vanishing order $\alpha $ under a chosen
risk-neutral measure and that $S$ cannot jump over the barrier $H$.
Then we can hedge the path dependent claim $X$ by the non-vanilla
European claim defined by $g$. The hedge works as follows
\begin{itemize}
\item If $X$ never knocks in, then the claim in~(\ref{eq:hedge-t})
expires worthless.

\item If and when the barrier is hit, we can
exchange~(\ref{eq:hedge-t}) for a claim on $f(S_{T})$ at zero costs,
as proved in~\cite{CL}.
\end{itemize}

If the asset process can jump over the barrier the hedging strategy
may no longer exactly replicate the knock-in claim. A criterion for
a superreplication in the self-dual case is given
in~\cite[Remark~5.17]{CL}; for quasi self-dual cases we refer
to~\cite[Remark~7.4]{MoS}. In practice the claim $g(S_T)$ could be
synthetically approximated by bonds, forwards and lots of vanilla
options, cf.~\cite{CL}.

The hedge in~(\ref{eq:hedge-t}) and also other hedges derived
in~\cite[Section~6.2]{CL} heavily depend on the order parameter
$\alpha$. The usual choice for $\alpha$ is then
$\alpha=1-2\lambda/\sigma_{BS}^{2}$ where $\lambda$ represents the
carrying costs and $\sigma_{BS}$ corresponds to the implied
volatility in the Black--Scholes model. But with this approach, the
empirically already quite well performing hedges (we again refer
to~\cite{eng:fen:nal:schw06,nal:pou06}) are {\it a priori}
Black--Scholes semi-static hedges. In view of that it seems natural
to discuss the derivation of semi-static hedging portfolios for
exponential L\'{e}vy processes in models with non-trivial carrying
costs. And exactly this will be the main application of the results
presented in the rest of this paper.

Related to the above discussion we should mention that different
asset price models can lead to the same hedge portfolios. Indeed,
if we assume that there are no-carrying costs then all assets
price models of the form $S=S_0\mathcal{E}(M)$, for $M$ being a
continuous Ocone martingale, would lead to the hedging
portfolio~(\ref{eq:hedge-t}) with $\alpha=1$. This can directly be
derived with the help of analogous conditioning arguments as
presented in~\cite{RhS}. A well-known example of an Ocone
martingale is
given by the solution of the stochastic differential equation%
\begin{align*}
dM_{t}  &  =V_{t}\,dB_{t},\\
dV_{t}  &  =-\mu V_{t}\,dt+\sqrt{V_{t}}\,dW_{t},
\end{align*}
where $\mu>0$ and $B$, $W$ are two independent Brownian motions,
see e.g.~\cite{RhS}. Further examples of Ocone martingales can
e.g.\ be found in~\cite{VY}.

\section{Exponential L\'{e}vy processes: general results}

\label{sec:gen-levy}

In this and the following section we consider a process $S$ which
is the exponential of a L\'{e}vy process $X$, $X_{0}=0$,
characterized by the
L\'{e}vy-Khintchine formula for the characteristic function $E(e^{iuX_{t}%
})=e^{t\psi(u)}$ for $u\in\mathbb{R}$ with
\begin{equation}
\psi(u)=\kappa(iu)\,,\quad\kappa(z)=\gamma z+\frac{1}{2}\sigma^{2}z^{2}%
+\int(e^{zx}-1-zxc(x))\nu(dx)\,, \label{eq:le-k}%
\end{equation}
with $c(x)=1\hspace*{-0.55ex}\mathrm{I}_{|x|\leq1}$, for
$z\in\mathbb{C}$, such that $\Re z=c$ satisfies
$\int_{|x|>1}e^{cx}\nu(dx)<\infty$ (the latter is the case if and
only if $Ee^{cX_{t}}<\infty$ for some $t>0$ or, equivalently, for
every $t>0$), where $\nu$ is the L\'{e}vy measure, i.e.\
$\nu(\{0\})=0$ and $\int\min(x^{2},1)\,\nu(dx)<\infty$, see
e.g.~\cite{Sat}, in particular Theorem~25.17.

There are no problems with strict local martingales here since a
local martingale which is a L\'{e}vy process is a martingale and
since a local martingale of the form $e^X$, with $X$ being a
L\'evy process, is a martingale, see e.g.~\cite[Lemma~4.4]{kal00}.
Moreover, if $Y$ is a L\'{e}vy martingale with $\Delta Y>-1$, then
$\mathcal{E(}Y\mathcal{)}$ is a martingale,
see~\cite[Corollary~7]{PS}.

Furthermore, as in~\cite{MoS}, we can use the L\'{e}vy property in
order to reduce the analysis of quasi self-duality to the analysis
of infinitely divisible distributions.

\begin{proposition}
\label{pro:rv}  Let $L$ be a L\'evy process with $L_{0}=0$.

\begin{itemize}

\item[$\mathrm{(A)}$] $L$ with $L_{0}=0$ is conditionally
symmetric if and only if  for any non-negative Borel function $f$
\[
E_{P}[f(L_{T})]=E_{P}[f(-L_{T})]\,.
\]

\item[$\mathrm{(B)}$] A process $S=S_{0}\exp(\lambda t+L)$ with
existing $\alpha$-moment for all $t\in[0,T]$ (or equivalently for
one $t>0$) is quasi self-dual  of order $\alpha$ if and only if
for any non-negative Borel function $f$ it holds that
\[
E_{P}\left[  f\left(  \frac{S_{T}}{S_{0}}\right)  \right]
=E_{P}\left[
\left(  \frac{S_{T}}{S_{0}}\right)  ^{\alpha}f\left(  \frac{S_{0}}{S_{T}%
}\right)  \right]  .
\]

\end{itemize}
\end{proposition}

An immediate consequence of~$\mathrm{(A)}$ is that an integrable
symmetric L\'{e}vy process is a martingale and in fact also a
conditionally symmetric one. Both observations are not true in
general cases, where one has to distinguish carefully between
different notions of symmetry, see e.g.~\cite{RhS} and the
literature cited therein. The fact that an integrable symmetric
L\'{e}vy process is a martingale is also a direct consequence
of~\cite[Exercise~18.1,~Example~25.12]{Sat} combined with~\cite[Propositions~3.17]%
{con:tan}.

\begin{proof}[Proof of Proposition~\ref{pro:rv}] $\mathrm{(A)}$ Symmetry of the distribution of
$L_{t}$ is not a time-dependent distributional property of
L\'{e}vy processes, see e.g.~\cite[Section~23]{Sat}. Furthermore,
by~\cite[Chapter~1,~Propoposition~6]{Ber} we have that for a
stopping time $\tau\in\lbrack0,T]$ the process $\tilde{L}$ defined
by $\tilde{L}_{s}=L_{\tau+s}-L_{\tau}$ is a copy of $L$
independent of $\mathcal{F}_{\tau}$ with
$\tilde{L}_{T-\tau}=L_{T}-L_{\tau}$. Hence,
\[
E[f(L_{T}-L_{\tau})|\mathcal{F}_{\tau}]=E[f(\tilde{L}_{T-\tau})|\mathcal{F}%
_{\tau}]=E[f(-\tilde{L}_{T-\tau})|\mathcal{F}_{\tau}]=E[f(L_{\tau}%
-L_{T})|\mathcal{F}_{\tau}]\,.
\]
$\mathrm{(B)}$~For non-vanishing $\alpha$ we can e.g.\ directly
combine the well-known results about L{\'e}vy processes
in~\cite[Theorems~7.10,~11.5,~Corollary~8.3,]{Sat} with the result
presented in~\cite[Theorem~5.3]{MoS} in order to see
that~$\mathrm{(B)}$ for $T>0$ already implies~$\mathrm{(B)}$ for all
other $t$. Furthermore, as a consequence of~$\mathrm{(A)}$, the same
is true for vanishing $\alpha$. The rest of the proof uses the above
argument.
\end{proof}

\medskip

In the light of Proposition~\ref{pro:rv} for $\alpha=1$, i.e.\ in
the put-call symmetry or self-dual case, the well-known result
in~\cite{faj:mor06} reads as follows, see also~\cite{CL} for the
conditional statement.
\begin{theorem}[\cite{faj:mor06,CL}]
\label{geometric PCS-g} Let $S=\exp(X)$ be a martingale for a
L\'{e}vy process $X$ with triplet $\left(
\gamma,\sigma^{2},\nu\right)  $
where $\nu$ is the L\'{e}vy measure. Then $S$ is self-dual iff%
\begin{equation}
\nu\left(  dx\right)  =e^{-x}\,\nu\left(  -dx\right)  .
\label{Levy symmetry-g}%
\end{equation}

\end{theorem}

\begin{remark}
\label{re:quasi-and-martingale-g} We stress that for self-duality
the martingale assumption is important in the above statement,
since for integrable $S=\exp(X)$~$(\ref{Levy symmetry-g})$ does
not imply self-duality. However, an integrable $S$ is self-dual if
and only if~$(\ref{Levy symmetry-g})$ holds along with the
``drift'' restriction which forces an integrable $S$ to be a
martingale, see~$\cite{MoS}$.
\end{remark}

In view of Proposition~\ref{characterization of quasi
self-duality} and Remark~\ref{re:quasi-and-martingale-g} it is
obvious that in the integrable quasi self-dual case of order
$\alpha\neq0$, $S^{\alpha}$ needs to be a martingale due to
symmetry reasons where $e^{-\lambda t}S$ is assumed to be a
martingale in order to obtain a risk-neutral setting. Hence, we
have only one ``drift'' which has to satisfy two (different)
martingale assumptions simultaneously, one for symmetry reasons
and the other one due to risk-neutrality. Roughly speaking we can
say that this intuitively describes the origin of~(\ref{relation
lambda - alpha-g}) in Theorem~\ref{quasi PCS-g}. Note that these
two conditions coincide in the self-dual case with vanishing
carrying costs.

The Conditions~$\mathrm{(i)}$,~$\mathrm{(ii)'}$,
and~$\mathrm{(iii)'}$ as well as,~$\mathrm{(a)}$
and~$\mathrm{(b)}$, respectively, will be unified in
Theorem~\ref{quasi PCS-g} to the
Conditions~$\mathrm{(i)}$,~$\mathrm{(ii)}$, and~$\mathrm{(iii)}$.

\begin{proposition}
\label{co:quasi PCS-g} Let $S=e^{\lambda t}\exp(X)$ for $\lambda\in%
\mathbb{R}
$ and a L\'{e}vy-process $X$ such that $S_{t}$ and
$(S_{t})^{\alpha}$ are integrable for some $t>0$. Then $S$ is
quasi self-dual of order $\alpha$ with $\exp(X)$ being a
martingale if and only if the following conditions hold.

\medskip

For $\alpha\neq0$:
\begin{description}
\item[$\mathrm{(i)}$] The L\'{e}vy measure satisfies
\begin{equation}
\nu\left(  dx\right)  =e^{-\alpha x}\,\nu\left(  -dx\right)  \,,
\label{Levy quasi self-dual-g}%
\end{equation}
i.e.\ $\nu(B)=\int_{-B}e^{\alpha x}\,d\nu(x)$ for a Borel set $B$
in $\mathbb{R}\setminus\{0\}$. \item[$\mathrm{(ii)'}$] The process
$S^{\alpha}$ is a martingale.

\item[$\mathrm{(iii)'}$] The process $\exp(X)$ is a martingale.
\end{description}

For $\alpha=0$:

\begin{itemize}
\item[$\mathrm{(a)}$] The process $X=\log(S)$ is a (conditionally)
symmetric L\'{e}vy process.

\item[$\mathrm{(b)}$] The process $S=\exp(X)$ is a martingale.
\end{itemize}
\end{proposition}

\begin{remark}
\label{re:mart-mart-sym=ok-g} Hence, if for $\alpha\neq0$
$S^{\alpha}$ and $\exp(X)$ are martingales (then the imposed
integrability assumptions are automatically satisfied) and
if~$\mathrm{(i)}$ holds, then we are in a risk-neutral setting
where the quasi self-duality
property~$(\ref{eq:def-quasi-equation})$ holds.

Note that as a consequence of~\cite[Proposition~11.10]{Sat}, the
L\'{e}vy triplet $(\tilde\gamma,\tilde\sigma^2,\tilde\nu)$ of
$Z=\alpha\lambda t+\alpha X$, $\alpha\neq0$, in terms of the triplet
$(\gamma,\sigma^{2},\nu)$ of $X$ is given by
\[
(\tilde{\gamma},\alpha^{2}\sigma^{2},(\nu\alpha^{-1}))\,,\quad\text{where}%
\quad\tilde{\gamma}=\alpha((\lambda+\gamma)+\int x(1\hspace*{-0.55ex}%
\mathrm{I}_{|\alpha
x|\leq1}-1\hspace*{-0.55ex}\mathrm{I}_{|x|\leq1})\nu(dx))
\]
and $(\nu\alpha^{-1})(B)=\nu(\{x\in\mathbb{R}:\alpha x\in B\})$.

\end{remark}

\begin{proof}[Proof of Proposition~\ref{co:quasi PCS-g}]
  For $\alpha\neq 0$ we
 only need to prove the equivalence of the Definition~\ref{def:qsd} with~$\mathrm{(i)}$
  and~$\mathrm{(ii)'}$, since $e^X$ is a martingale in either case. If $S$ is quasi self-dual of order $\alpha$ then we can apply
 Definition~\ref{eq:def-quasi-equation} for $f$ being identically
 one and $\tau=0$ in order to see that given the imposed integrability assumptions
 $E[e^{Z_T}]=1$, i.e.\
 $S^\alpha$ is a martingale, since $Z$ is a L\'evy process,
 see~\cite[Proposition~3.17]{con:tan}. Furthermore, Proposition~\ref{characterization of quasi
 self-duality} then implies that $e^Z$ is self-dual so that by
 Theorem~\ref{geometric PCS-g} the L\'evy measure $\tilde\nu$ satisfies~(\ref{Levy
 symmetry-g}). Hence, for any Borel set $B\subset\mathbb R\setminus\{0\}$
 \begin{multline*}
   \int_B \nu(dx)=\int
   1\hspace*{-0.55ex}\mathrm{I}_{B}(\alpha^{-1}y)\tilde\nu(dy)\\
   =\int
   1\hspace*{-0.55ex}\mathrm{I}_{B}(\alpha^{-1}y)e^{-y}\tilde\nu(-dy)
   =\int 1\hspace*{-0.55ex}\mathrm{I}_{B}(-\alpha^{-1}y)e^{y}\tilde\nu(dy)
   =\int_{-B} e^{\alpha x}\nu(dx)\,,
 \end{multline*}
 i.e.~(\ref{Levy quasi self-dual-g}) holds so that~$\mathrm{(i)}$ and~$\mathrm{(ii)'}$ hold.

 \medskip

 Conversely, given~$\mathrm{(ii)'}$ from Proposition~\ref{co:quasi
 PCS-g} we have for every Borel set $B\subset \mathbb R\setminus
 \{0\}$
 \begin{align*}
   \int_B \tilde \nu(dy)&=\int 1\hspace*{-0.55ex}\mathrm{I}_{ B}(\alpha x)\nu(dx)
   =\int 1\hspace*{-0.55ex}\mathrm{I}_{B}(\alpha x)e^{-\alpha x}\nu(-dx)
   =\int 1\hspace*{-0.55ex}\mathrm{I}_{ B}(-\alpha x)e^{\alpha x}\nu(dx)
   =\int_{-B}e^{y}\tilde\nu(dy)\,,
 \end{align*}
 so that~(\ref{Levy symmetry-g}) follows. Since furthermore $e^Z$ is now assumed to be
 a martingale, we obtain its self-duality property by
 Theorem~\ref{geometric PCS-g}. By Proposition~\ref{characterization of quasi
 self-duality} we end up with the quasi self-duality of
 $S$.

  For $\alpha=0$ we can use again the strong Markov as well as the
  independent and stationarity property of the increments of
  L\'{e}vy processes in order to see
  that~(\ref{eq:def-quasi-equation}) is equivalent to the property
  that $Z_{t}=(\lambda t+X_{t})$ has an even distribution for some,
  or equivalently, for all $t\in(0,T]$ (recall that symmetry is not
  a time-dependent distributional property of a L\'{e}vy process)
  since the condition is reflected in the L\'evy triplet characterizing
  the distribution of the whole process.
\end{proof}

\medskip

For the case $\alpha\neq0$ the following theorem is a summary of
various results presented in~\cite{MoS}, adapted to the present
setting, with an alternative proof based on
Proposition~\ref{co:quasi PCS-g}. In order to exclude any confusion
related to carrying costs let us stress that consistently
with~(\ref{eq:le-k}) the meaning of the L\'evy triplet of a L\'evy
process $X$ is that $P_{X_1}=\mu$ for an infinitely divisible
distribution $\mu$ and $P_{X_t}=\mu^t$, where the latter infinitely
divisible \emph{distributions} have generating triplet
$(t\gamma,tA,t\nu)$ (we assume w.l.o.g.\ that the process is defined
on an interval including $1$).

\begin{theorem}
\label{quasi PCS-g} Let $S=e^{\lambda t}\exp(X)$ for $\lambda\in%
\mathbb{R}
$ and a L\'{e}vy-process $X$ with triplet $\left(  \gamma,\sigma^{2}%
,\nu\right)  $, such that $S_{t}$ and $(S_{t})^{\alpha}$ are
integrable for some $t>0$. Then $S$ is quasi self-dual of order
$\alpha$ with $\exp(X)$ being a martingale if and only if the
following conditions hold.

\begin{description}
\item[$\mathrm{(i)}$] The L\'{e}vy measure satisfies
Condition~$\mathrm{(i)}$ from Proposition~$\ref{co:quasi PCS-g}$.
 \item[$\mathrm{(ii)}$] The entries of the triplet satisfy
\begin{equation}
\gamma=\int_{|x|\leq1}x\left(  1-e^{\frac{1}{2}\alpha x}\right)
\,\nu\left( dx\right)  -\frac{1}{2}\alpha\sigma^{2}-\lambda.
\label{Esscher martingale condition-g}%
\end{equation}

\item[$\mathrm{(iii)}$] The parameters $\lambda$ and $\alpha$ are related by%
\begin{equation}
\lambda=\left(  1-\alpha\right)  \frac{\sigma^{2}}{2}+\int\left(
e^{x}-xe^{\frac{1}{2}\alpha
x}1\hspace*{-0.55ex}\mathrm{I}_{|x|\leq
1}-1\right)  \,\nu\left(  dx\right)  . \label{relation lambda - alpha-g}%
\end{equation}

\end{description}
\end{theorem}

\begin{proof}
Given the Conditions $\mathrm{(i)}$, $\mathrm{(ii)}$, and
$\mathrm{(iii)}$ and the imposed integrability assumptions we can
substitute~(\ref{relation lambda - alpha-g}) in~(\ref{Esscher
martingale condition-g}) in order to obtain
 \begin{equation}
\gamma=-\frac{\sigma^{2}}{2}+\int\left(  x1\hspace*{-0.55ex}\mathrm{I}%
_{|x|\leq1}+1-e^{x}\right)  \,\nu\left(  dx\right)  \,,
\label{eq:mart-drift-g}\
\end{equation}
so that in view of the imposed integrability
assumptions~$\mathrm{(iii)'}$ follows, see
e.g.~\cite[Proposition~3.18]{con:tan}. Furthermore, by
multiplying~(\ref{Esscher martingale condition-g}) with
$\alpha\neq0$, adding $\alpha\int
x(1\hspace*{-0.55ex}\mathrm{I}_{|\alpha x|\leq1}-1\hspace
*{-0.55ex}\mathrm{I}_{|x|\leq1})\nu(dx)$,
Condition~$\mathrm{(ii)}$ can be written as
\begin{equation}
\alpha(\gamma+\lambda+\int x(1\hspace*{-0.55ex}\mathrm{I}_{|\alpha
x|\leq
1}-1\hspace*{-0.55ex}\mathrm{I}_{|x|\leq1})\nu(dx))=-\frac{1}{2}(\alpha
\sigma)^{2}-\alpha\int_{\mathbb{R}}x(e^{\frac{1}{2}\alpha
x}1\hspace
*{-0.55ex}\mathrm{I}_{|x|\leq1}-1\hspace*{-0.55ex}\mathrm{I}_{|\alpha
x|\leq
1})\nu(dx)=\tilde\gamma\,. \label{eq:mult-alpha-g}%
\end{equation}
On the other hand, by substituting, we can rewrite the integral
expression in the (given the integrability) martingale condition
of $S^{\alpha}$, i.e.
\begin{equation}
\tilde{\gamma}=-\frac{1}{2}(\alpha\sigma)^{2}+\int(x1\hspace*{-0.55ex}%
\mathrm{I}_{|x|\leq1}+1-e^{x})(\nu\alpha^{-1})(dx)\,,
\label{eq:power-mart-cond-g}%
\end{equation}
as
\[
\int(e^{y}-1-y1\hspace*{-0.55ex}\mathrm{I}_{|y|\leq1})(\nu\alpha
^{-1})(dy)=\int(e^{\alpha x}-1-\alpha
x1\hspace*{-0.55ex}\mathrm{I}_{|\alpha x|\leq1})\,\nu(dx)\,.
\]
Hence,~(\ref{eq:mult-alpha-g}) coincides
with~(\ref{eq:power-mart-cond-g}) since $\int(e^{\alpha
x}-1-\alpha xe^{\frac{1}{2}\alpha x}1\hspace
*{-0.55ex}\mathrm{I}_{|x|\leq1})\nu(dx)$ vanishes. The latter is a
consequence of~$\mathrm{(i)}$, concretely
\begin{multline}
\int(e^{\alpha x}-1-\alpha xe^{\frac{1}{2}\alpha x}1\hspace*{-0.55ex}%
\mathrm{I}_{|x|\leq1})\nu(dx)\\
=\int_{-\infty}^{0}(e^{\alpha x}-1-\alpha xe^{\frac{1}{2}\alpha x}%
1\hspace*{-0.55ex}\mathrm{I}_{|x|\leq1})e^{-\alpha x}\nu(-dx)+\int_{0}%
^{\infty}(e^{\alpha x}-1-\alpha xe^{\frac{1}{2}\alpha x}1\hspace
*{-0.55ex}\mathrm{I}_{|x|\leq1})\nu(dx)\\
=\int_{0}^{\infty}(1-e^{\alpha x}+\alpha xe^{\frac{1}{2}\alpha x}%
1\hspace*{-0.55ex}\mathrm{I}_{|x|\leq1})\,\nu(dx)\\
+\int_{0}^{\infty}(e^{\alpha x}-1-\alpha xe^{\frac{1}{2}\alpha x}%
1\hspace*{-0.55ex}\mathrm{I}_{|x|\leq1})\nu(dx)=0\,.
\label{eq:vanishing-int-g}%
\end{multline}

\medskip

Conversely, if we start with $\mathrm{(i)}$, $\mathrm{(ii)'}$, and
$\mathrm{(iii)'}$ we can use~$\mathrm{(i)}$ to see
that~(\ref{eq:vanishing-int-g}) holds again so
that~$\mathrm{(ii)'}$ can be rewritten as~$\mathrm{(ii)}$
(respectively,~(\ref{eq:power-mart-cond-g}) as~(\ref{Esscher
martingale condition-g})) by the converse calculations from above
(for $\alpha\neq0$). By equating the martingale condition $
\gamma=-\frac{\sigma^{2}}{2}+\int\left(  x1\hspace*{-0.55ex}\mathrm{I}%
_{|x|\leq1}+1-e^{x}\right)  \,\nu\left(  dx\right)
$ with~(\ref{Esscher martingale condition-g}) we arrive
at~(\ref{relation lambda - alpha-g}), i.e.\
Condition~$\mathrm{(iii)}$ is implied. Essentially the same proof
results if we use that the martingale condition of
$e^{\alpha(\lambda t+X)}$ is equivalent to $E[e^{\alpha(\lambda
t+X_t)}]=1$ and then apply~\cite[Theorem~27.15]{Sat} instead
of~\cite[Proposition~11.10]{Sat}.

\medskip

For $\alpha=0$ recall that we can use the strong Markov as well as
the independent and stationarity property of the increments of
L\'{e}vy processes in order to see
that~(\ref{eq:def-quasi-equation}) is equivalent to the property
that $Z_{t}=(\lambda t+X_{t})$ has an even distribution for some,
or equivalently, for all $t\in(0,T]$ (where symmetry is not a
time-dependent distributional property of a L\'{e}vy process),
since the latter is the case if and only if the triplet
$(\gamma+\lambda,\sigma ^{2},\nu)$ of $Z$ satisfies that
$\gamma+\lambda$ vanishes and $\nu$ is even, see again~
\cite[Exercise~18.1]{Sat}, i.e.\ if and only if~$\mathrm{(i)}$
and~$\mathrm{(ii)}$ are satisfied for $\alpha=0$. In view of the
existence of the first exponential moment we have that the
martingale property of $\exp(X)$ is equivalent
to~(\ref{eq:mart-drift-g}). Hence, quasi self-duality of order
$\alpha=0$ implies~$\mathrm{(i)}$ and~$\mathrm{(ii)}$ for
vanishing $\alpha$ and since furthermore $\exp(X)$ is a martingale
we can plug~$\mathrm{(ii)}$ for vanishing $\alpha$
in~(\ref{eq:mart-drift-g}) in order to end up
with~$\mathrm{(iii)}$ for vanishing $\alpha$.
Conversely,~$\mathrm{(i)}$ and~$\mathrm{(ii)}$ for vanishing
$\alpha$ imply quasi self-duality of order $\alpha=0$ and by
plugging~$\mathrm{(ii)}$ in~$\mathrm{(iii)}$ for vanishing
$\alpha$ we end up with~(\ref{eq:mart-drift-g}) implying the
martingale property of $\exp(X)$, given the assumed integrability
of $\exp(X)$.
\end{proof}

\medskip

\begin{remark}
\label{re:merits} Both formulations, the one from
Proposition~$\ref{co:quasi PCS-g}$ and the one from
Theorem~$\ref{quasi PCS-g}$ have their merits. An advantage of the
first formulation is that this statement does not depend anymore
on the choice of the function $c(\cdot)$ in $\psi$,
while~$(\ref{relation lambda -
 alpha-g})$ will be particularly useful for uniqueness discussions
 of the parameter $\alpha$. Furthermore, the formulation
 in Theorem~$\ref{quasi PCS-g}$ needs no distinction of the case of vanishing $\alpha$.
 For the concrete derivation of $\alpha$, both characterizations can be helpful.
\end{remark}

Now we discuss an analogue of Theorem~\ref{quasi PCS-g} for the
case when the price process is represented as a stochastic rather
than an ordinary exponential. As preparation we recall the
following well-known result which can be found in~\cite{JS},
Theorem II.8.10.

\begin{proposition}
\label{Levy equivalence} \textbf{Equivalence between two different
representations of exponential L\'{e}vy processes. } Let $X$, $Y$
be two L\'{e}vy processes such that $X_{0}=0$ and $\Delta Y>-1$.
Then we have
$\exp(X)=\mathcal{E}\left(  Y\right)  $ if and only if%
\begin{align}
X  &  =Y-Y_{0}-\frac{1}{2}\left[  Y^{c}\right]  +\left( \log\left(
1+y\right)  -y\right)  \ast\mu^{Y},\label{equivalence relations}\\
Y  &  =Y_{0}+X+\frac{1}{2}\left[  X^{c}\right]  +\left(
e^{x}-1-x\right)
\ast\mu^{X},\nonumber\\
\qquad x\ast\mu^{X}  &  =\log\left(  1+y\right) \ast\mu^{Y},\qquad
y\ast \mu^{Y}=\left(  e^{x}-1\right) \ast\mu^{X}.\nonumber
\end{align}
Here $\mu^{X}$, $\mu^{Y}$ are the jump measures of $X$,
respectively $Y$.
\end{proposition}

In the following we denote by $\left(
\gamma^{X},\sigma^{2},\nu^{X}\right) $, $\left(
\gamma^{Y},\sigma^{2},\nu^{Y}\right)  $ the triplets of $X$,
respectively $Y$ and we again put $X_0=0$, $Y_0=0$. The relations
between the L\'evy triplets are e.g.\ given
in~\cite[Corollary~4.1]{BN:SHY}. In the sequel we apply the
self-inverse function $\chi: (-1,\infty )\to(-1,\infty)$, defined
by
\[
\chi(y)=-\frac{y}{1+y}\,.
\]

Note that by setting $\alpha=1$ and $\lambda=0$, we obtain an
analogon to the characterization of self-dual continuous processes
obtained by Tehranchi in~\cite[Theorem~3.1]{T}. However, the
characterization in terms of the conditional symmetry of the
stochastic logarithm does not hold in this L\'{e}vy setting, as in
particular we here have to invoke the (restricted) M\"{o}bius
transform $\chi$.

\begin{theorem}
\label{pr:tehr-levy} In the setting of Theorem~$\ref{quasi PCS-g}$,
for $\alpha\neq0$ let $Y$, $\tilde{Y}$ be L\'{e}vy processes with
$\Delta Y\,,\Delta\tilde{Y}>-1$ related to $X$ and $Z=\alpha(\lambda
t+X)$ by~$(\ref{equivalence relations})$, respectively (i.e.\
$\mathcal{E}(Y)=\exp(X)$ and $\mathcal{E}(\tilde{Y})=S^{\alpha}$).
Then $S$ is quasi self-dual of order $\alpha$ with
$\mathcal{E}\left(  Y\right)  $ being a martingale if and only if
the following conditions hold.

\begin{description}
\item[$\mathrm{(a)}$] For all Borel sets $B$ the L\'{e}vy measure
satisfies
\[
\nu^{Y}(B) =\int_{\chi(B)}(1+y)^{\alpha}\nu^{Y}(dy)\,,
\]

\item[$\mathrm{(b)}$] The process $\tilde Y$ is a martingale.

\item[$\mathrm{(c)}$] The process $Y$ is a martingale.
\end{description}
\end{theorem}

Note that Condition~$\mathrm{(a)}$ can equivalently be written as
\begin{equation}
\label{eq:mellin}\nu^{Y}(B)=\int_{B}(1+y)^{-\alpha}\left(
\nu^{Y}\chi ^{-1}\right)  (dy)\,.
\end{equation}

\begin{proof}
Since $\mathcal{E}(Y)=\exp(X)$ we have to show the equivalence of
Conditions~$\mathrm{(a)}$, $\mathrm{(b)}$, and $\mathrm{(c)}$ with
the Condition~$\mathrm{(i)}$ from Theorem~\ref{quasi PCS-g} and
the Conditions~$\mathrm{(ii)'}$ as well as~$\mathrm{(iii)'}$ from
Proposition~\ref{co:quasi PCS-g}.

 If~(\ref{Levy
quasi self-dual-g}) holds then
\begin{align*}
\nu^{Y}\left(  B\right)   &
=\int1\hspace*{-0.55ex}\mathrm{I}_{B}(y)\nu ^{Y}\left(  dy\right)
=\int1\hspace*{-0.55ex}\mathrm{I}_{B}(e^{x}-1)\nu
^{X}\left(  dx\right) \\
&  =\int1\hspace*{-0.55ex}\mathrm{I}_{B}(e^{x}-1)e^{-\alpha x}\nu
^{X}(-dx)=\int1\hspace*{-0.55ex}\mathrm{I}_{B}(e^{-x}-1)e^{\alpha
x}\nu
^{X}(dx)\\
&  =\int1\hspace*{-0.55ex}\mathrm{I}_{B}\left(
\frac{-y}{1+y}\right) (1+y)^{\alpha}\nu^{Y}\left(  dy\right)
=\int_{\chi(B)}(1+y)^{\alpha}\nu ^{Y}\left(  dy\right)  \,,
\end{align*}
where on the r.h.s.\ of the last equation we can equivalently
write $\int _{B}(1+y)^{-\alpha}\left(  \nu^{Y}\chi^{-1}\right)
(dy)$, i.e.\ Condition~$\mathrm{(a)}$ follows. If~$\mathrm{(a)}$
holds then
\begin{align*}
\nu^{X}\left(  B\right)   &
=\int1\hspace*{-0.55ex}\mathrm{I}_{B}(x)\nu ^{X}\left(  dx\right)
=\int1\hspace*{-0.55ex}\mathrm{I}_{B}(\log(1+y))\nu
^{Y}\left(  dy\right) \\
&
=\int1\hspace*{-0.55ex}\mathrm{I}_{B}(\log(1+y))(1+y)^{-\alpha}\left(
\nu^{Y}\chi^{-1}\right)  \left(  dy\right)  =\int1\hspace*{-0.55ex}%
\mathrm{I}_{B}\left(  \log\left(  \frac{1}{1+y}\right)  \right)
(1+y)^{\alpha}\nu^{Y}\left(  dy\right) \\
&  =\int1\hspace*{-0.55ex}\mathrm{I}_{B}\left(  -x\right)
e^{\alpha x}\nu ^{X}\left(  dx\right)  =\int_{-B}e^{\alpha
x}\nu^{X}\left(  dx\right)  \,,
\end{align*}
i.e.\ we end up with~(\ref{Levy quasi self-dual-g}).

The remaining two equivalences are a consequence of the L\'evy
property, since the stochastic exponential $\mathcal{E}(Z)$ for a
L\'evy process $Z$ is a (local) martingale iff so is $Z$.
\end{proof}

\medskip

\begin{remark}
\ If we want to include the special case of vanishing $\alpha$ in
the context of Theorem~$\ref{pr:tehr-levy}$ then we can rewrite
the conditions of Theorem~$\ref{quasi PCS-g}$. As in the proof of
Theorem~$\ref{pr:tehr-levy}$ we then end up with
Condition~$\mathrm{(a)}$ for the L\'evy measure. The other two
conditions are

\begin{itemize}
\item[$\mathrm{(b')}$]
$\gamma^{Y}=\frac{1}{2}\sigma^{2}(1-\alpha)-\lambda+\int
(y1\hspace*{-0.55ex}\mathrm{I}_{|y|\leq1}-\log(1+y)(1+y)^{\frac{1}{2}\alpha
}1\hspace*{-0.55ex}\mathrm{I}_{|\log(1+y)|\leq1})\nu^{Y}\left(
dy\right)  $,

\item[$\mathrm{(c')}$]
$\lambda=\frac{1}{2}\sigma^{2}(1-\alpha)+\int(y-\log
(1+y)(1+y)^{\frac{1}{2}\alpha}1\hspace*{-0.55ex}\mathrm{I}_{|\log(1+y)|\leq
1})\nu^{Y}\left(  dy\right)  $.
\end{itemize}

Condition~$\mathrm{(b')}$ is obtained by changing variables and
applying~$\cite[Corollary~4.1]{BN:SHY}$ while~$\mathrm{(c')}$ is
simply obtained by changing variables. Note that given that $Y$ is
integrable plugging in~$\mathrm{(c')}$ into~$\mathrm{(b')}$ yields
the martingale condition for $Y$.
\end{remark}

The next result excludes certain combinations of parameters.

\begin{proposition}
\label{prop:jensen} Assume that $\exp(X)$ is a martingale,
$S=e^{\lambda t}\exp(X)$ has finite $\alpha$-moments, and either

\begin{itemize}
\item $\lambda>0$ along with $\alpha>1$\,,

\item $\lambda<0$ along with $\alpha<1$, $\alpha\neq0$,
\end{itemize}

Then $S$ cannot be quasi self-dual of order $\alpha$.
\end{proposition}

Note that the omitted case $\alpha=1$ corresponds to the self-dual
case where $\lambda$ needs to vanish (related to a self-dual
L\'evy measure) and
$\alpha=0$ where $\gamma=-\lambda=-\frac{1}{2}\sigma^{2}-\int(e^{x}%
-1-x1\hspace*{-0.55ex}\mathrm{I}_{|x|\leq1})\nu_{0}\left(
dx\right)  $ is needed in order to end up with the corresponding
quasi self-duality along with risk neutrality.

\begin{proof}
For $\alpha>1$, $f:(0,\infty)\rightarrow(0,\infty)$,
$f(x)=x^{\alpha}$ is strictly convex increasing so that by the
Jensen inequality, the martingale property of $\exp(X)$ and the
assumption $\lambda>0$ we obtain for $t>s$ that
\[
E\big[\big(e^{\lambda
t+X_{t}}\big)^{\alpha}|\mathcal{F}_{s}\big]\geq\left(
E\big[e^{\lambda t+X_{t}}|\mathcal{F}_{s}\big]\right)
^{\alpha}=\left( e^{\lambda t}e^{X_{s}}\right)  ^{\alpha}>\left(
e^{\lambda s+X_{s}}\right) ^{\alpha}\,,
\]
i.e.\ $S^{\alpha}$ cannot be a martingale so that Condition
$\mathrm{(ii)'}$ of Proposition~\ref{co:quasi PCS-g} does not
hold.
\medskip

For $\alpha\in(0,1)$ the function $-f$ is strictly convex and
increasing so that the Jensen inequality, the martingale property
of $\exp(X)$ and $\lambda<0$ imply that $S^{\alpha}$ is a
supermartingale, but not a martingale, so that the quasi
self-duality of order $\alpha$ can again not hold. \medskip

The last case follows again by the fact that for $\alpha<0$ the
function $f$ is strictly convex but now decreasing.
\end{proof}

\medskip

In view of Proposition~\ref{prop:jensen} we now turn our attention
to the integral in Equation~(\ref{relation lambda - alpha-g}), and
discuss first an important monotonicity property.

\begin{proposition}
\label{pr:mon} Assume that the L\'{e}vy measure can be written in
the form
\begin{equation}
\nu(dx)=e^{-\frac{1}{2}\alpha x}\nu_{0}(dx)\,, \label{eq:nice-form}%
\end{equation}
with an even measure $\nu_{0}$, $\nu_{0}(dx)=\nu_{0}(-dx)$, being
independent of $\alpha$. Furthermore, take $\alpha_{1}<\alpha_{2}$
and assume that $E\left[  S_{t}^{\alpha_{i}}\right]  <\infty$ for
$\alpha_{i}=\alpha _{1}\,,\alpha_{2}\,,1$, and some $t>0$. Then
the integral in~$(\ref{relation lambda - alpha-g})$ satisfies
\begin{equation}
\int\left(  e^{x}-xe^{\frac{\alpha_{1}}{2}x}1\hspace*{-0.55ex}\mathrm{I}%
_{|x|\leq1}-1\right)  \,\nu\left(  dx\right)  \geq\int\left(  e^{x}%
-xe^{\frac{\alpha_{2}}{2}x}1\hspace*{-0.55ex}\mathrm{I}_{|x|\leq1}-1\right)
\,\nu\left(  dx\right)  \,. \label{eq:mon}%
\end{equation}
If furthermore the measure $\nu_{0}$ satisfies that $\nu_{0}(B)>0$
for some Borel set in $\mathbb{R}\setminus\{0\}$ then the
inequality in~$(\ref{eq:mon})$ is strict.
\end{proposition}

As a consequence of H{\"o}lder's inequality the set of existing
$\alpha$-moments is convex, see~\cite[Theorem~25.17]{Sat}.

\begin{proof}
Assume that $\alpha_{1}<\alpha_{2}$. By assumption the integrals
in~(\ref{eq:mon}) can be written as
\[
\int\left(  e^{(1-\frac{1}{2}\alpha_{i})x}-x1\hspace*{-0.55ex}\mathrm{I}%
_{|x|\leq1}-e^{-\frac{\alpha_{i}}{2}x}\right)  \,\nu_{0}\left(
dx\right) \,,\quad i=1,2\,.
\]
If we consider for every \emph{fixed}
$x\in\mathbb{R}\setminus\{0\}$ the integrand as a function of
$\alpha$, i.e.
\[
\ g_{x}(\alpha)=e^{(1-\frac{1}{2}\alpha)x}-x1\hspace*{-0.55ex}\mathrm{I}%
_{|x|\leq1}-e^{-\frac{1}{2}\alpha x}\,,
\]
then we obtain a family of differentiable functions with
\[
g_{x}^{\prime}(\alpha)=\frac{1}{2}xe^{-\frac{1}{2}\alpha
x}(1-e^{x})\,.
\]
Since for every $x>0$ and any $\alpha\in\mathbb{R}$ (note that we
only consider the integrand, the integral does not need to
converge for any $\alpha\in\mathbb{R}$)
$\frac{1}{2}xe^{-\frac{1}{2}\alpha x}>0$ while $1-e^{x}<0$, and
because for every $x<0$ and any $\alpha\in\mathbb{R}$ we have that
$\frac{1}{2}xe^{-\frac{1}{2}\alpha x}<0$ along with $1-e^{x}>0$,
we end up with $g_{x}^{\prime}(\alpha)<0$, i.e.\ the functions
$g_{x}(\alpha)$ are
strictly monotonically decreasing. Hence, for every $x\in\mathbb{R}%
\setminus\{0\}$ we have that
\[
e^{(1-\frac{1}{2}\alpha_{1})x}-x1\hspace*{-0.55ex}\mathrm{I}_{|x|\leq
1}-e^{-\frac{\alpha_{1}}{2}x}>e^{(1-\frac{1}{2}\alpha_{2})x}-x1\hspace
*{-0.55ex}\mathrm{I}_{|x|\leq1}-e^{-\frac{\alpha_{2}}{2}x}\,.
\]
Since the integral is order preserving we end up with the claim.
\end{proof}

\medskip

We will see in Section~\ref{sec:Meixner} that the
monotonicity~(\ref{eq:mon}) does not always hold. However, as an
immediate consequence of Proposition~\ref{pr:mon}, the
monotonicity holds for the well-known family of Generalized
Hyperbolic (GH) models and also for the CGMY model.

\begin{remark}
[Monotonicity in Generalized Hyperbolic models]\label{re:GH-mon}
\textrm{The infinitely divisible Generalized Hyperbolic
distribution was introduced by Barndorff-Nielsen
in~$\cite{BN:77}$. The corresponding processes became quite
popular in the recent financial literature, see
e.g.~$\cite{BN:SHY,Sch}$ and the literature cited therein. GH
processes have no centered Gaussian term and their
L\'{e}vy measures have a density given by }%
\begin{equation}
\nu(x)=\frac{e^{bx}}{|x|}\left(
\int_{0}^{\infty}\frac{\exp(-|x|\sqrt
{2y+a^{2}})}{\pi^{2}y(J_{|v|}^{2}(d\sqrt{2y})+N_{|v|}^{2}(d\sqrt{2y}%
))}\,dy+\max(0,v)e^{-a|x|}\right)  \,, \label{eq:GH-nu}%
\end{equation}
\textrm{where the functions }$J_{v}$\textrm{ and }$N_{v}$\textrm{
are the Bessel functions of the first and second kind, see e.g.
$\cite[\mathrm{Appendix~A}]{Sch}$,
and where we consider the parameters restricted to }$d>0$\textrm{, }%
$a>\frac{1}{2}$\textrm{, }$-a<b<a-1$\textrm{, in order to ensure
that the first exponential moment always exists. If we rewrite the
density by }$b=-\frac{1}{2}\alpha$\textrm{ for
}$\alpha\in(-2(a-1),2a)$\textrm{, then the L\'{e}vy measure is of
the form~$(\ref{eq:nice-form})$ where the }$\alpha$\textrm{th
exponential moments also exist. Hence, with the imposed parameter
restrictions and with respect to the L\'{e}vy measure with density
given in~$(\ref{eq:GH-nu})$ with }$b=-\frac{1}{2}\alpha$\textrm{,
the integral in~$(\ref{relation lambda - alpha-g})$ strictly
decreases as a function of }$\alpha\in(-2(a-1),2a)$\textrm{.}
\end{remark}

\begin{remark}
[Monotonicity in the CGMY model]\label{re:CGMH-mon} \textrm{In the
classical four-parameter CGMY \linebreak model we again have no
centered Gaussian term and the
L\'{e}vy measure has a density given by }%
\begin{equation}
\nu(x)=\frac{C}{|x|^{1+Y}}e^{-G|x|}1\hspace*{-0.55ex}\mathrm{I}_{x<0}+\frac
{C}{|x|^{1+Y}}e^{-M|x|}1\hspace*{-0.55ex}\mathrm{I}_{x>0}\,,
\label{eq:nu-CGMY}%
\end{equation}
where we consider the parameters $C>0$\textrm{, }$G>0$\textrm{,
}$M>1$\textrm{, }$Y<2$ in order to ensure that the first
exponential moments exist. For $Y=0$\textrm{ we obtain VG
processes, which are considered in detail in
Section~\ref{subs:VG}. For negative }$Y$\textrm{ we obtain
compound Poisson models, for }$0<Y<1$\textquotedblleft infinite
activity\textquotedblright\ models (i.e.\ on any time interval
$(0,t]$, $t>0$, the process $X_{t}$, $t\geq0$, has, with
probability one, an infinite number of jumps) exhibiting
trajectories
of finite variation. If $Y\geq1$\textrm{ the variation is infinite. When }%
$Y$\textrm{ is close to }$2$\textrm{ then the process behaves much
like a Brownian motion, for more details we refer
e.g.~to~$\cite[\mathrm{Section~4.5}]{con:tan}$.}

\textrm{By choosing }%
\begin{equation}
M=\beta+\frac{1}{2}\alpha>1\,,\quad\text{ and}\quad G=\beta-\frac{1}{2}%
\alpha>0\,, \label{eq:CGMY-st-qusi-sd}%
\end{equation}
\textrm{we ensure that the L\'{e}vy measure is given by }%
\[
\nu(x)=e^{-\frac{1}{2}\alpha
x}\nu_{0}(x)\,,\quad\text{with}\quad\nu
_{0}(x)=\frac{Ce^{-\beta|x|}}{|x|^{1+Y}}1\hspace*{-0.55ex}\mathrm{I}_{x\neq
0}\,,
\]
\textrm{i.e.\ we have that Condition~$(\ref{eq:nice-form})$ is
satisfied so that the integral expression in~$(\ref{relation
lambda - alpha-g})$, respectively,
in~$\cite[\mathrm{Remark~5.8}]{MoS}$, is strictly monotonically
decreasing in }$\alpha \in(-2(\beta-1),2\beta)$\textrm{, where
}$\beta>\frac{1}{2}$\textrm{ is needed in order to ensure that
this interval is not empty. The integral expression
in~$\cite[\mathrm{Remark~5.8}]{MoS}$ is related to the choice of
$c(x)=1$ in $\psi$ being possible and quite popular in the CGMY
model. The proof of the monotonicity property in this case is
easily obtained by changing $c(x)=1\hspace
*{-0.55ex}\mathrm{I}_{|x|\leq1}$ to $c(x)=1$ in the proof of
Proposition~$\ref{pr:mon}$.}
\end{remark}

In order to efficiently derive a suitable $\alpha$ for given
carrying costs $\lambda$, the following result is often useful. We
stress that in the following proposition, $\psi_{\cdot}$,
$\kappa_{\cdot}$ are defined as in~(\ref{eq:le-k}) with
$c(x)=1\hspace*{-0.55ex}\mathrm{I}_{|x|\leq1}$.

\begin{proposition}
\label{prop:short-way-basis} Let $S=e^{\lambda t}\exp(X)$ for
$\lambda \in\mathbb{R}$ and a L\'{e}vy-process $X$ with triplet
$\left(  \gamma ,\sigma^{2},\nu\right)  $ (with corresponding
$\psi$), such that $S_{t}$ and $(S_{t})^{\alpha}$ are integrable
for some $t>0$. Furthermore, assume that the L\'{e}vy measure is
of the form
\begin{equation}
\nu(dx)=e^{-\frac{1}{2}\alpha x}\nu_{0}^{\alpha}(dx)
\label{eq:n0-dep-on-alpha}%
\end{equation}
with $\nu_{0}^{\alpha}(dx)=\nu_{0}^{\alpha}(-dx)$, i.e.\ even but
possibly depending on $\alpha$. Then Condition~$(\ref{relation
lambda - alpha-g})$ can be written as
\begin{multline}
\lambda=\left(  1-\alpha\right)  \frac{\sigma^{2}}{2}+\int\left(
e^{x}-xe^{\frac{1}{2}\alpha
x}1\hspace*{-0.55ex}\mathrm{I}_{|x|\leq
1}-1\right)  \,\nu\left(  dx\right) \label{eq:short-way}\\
=\psi_{0}^{(\alpha)}\big(-i(1-\frac{1}{2}\alpha)\big)-\psi_{0}^{(\alpha)}\big(i\frac{1}{2}%
\alpha\big)=\kappa_{0}^{(\alpha)}\big(1-\frac{1}{2}\alpha\big)-\kappa
_{0}^{(\alpha)}\big(-\frac{1}{2}\alpha\big)
\end{multline}
where $\psi_{0}$, $\kappa_{0}$, correspond to the triplet $\left(
0,\sigma^{2},\nu_{0}^{\alpha}\right)  $.
\end{proposition}

Hence, with the above observation the task of finding $\alpha$ for
given $\lambda$ is similar to the task of finding the parameter
for the Esscher martingale transform for exponential processes,
see~\cite{HS}. Note however, that $\psi_0^{(\alpha)}$,
$\kappa_0^{(\alpha)}$ may depend on $\alpha$, which can be a
source of non-uniqueness of $\alpha$ in
inverting~(\ref{eq:short-way}). For some restricted special cases
where the L\'evy process has vanishing centered Gaussian part and
the L\'evy measure $\nu_0^\alpha$ has finite Laplace transform on
the real line, a corresponding result in terms of the Laplace
transform of the L\'evy measure has been derived
in~\cite[Remark~5.8]{MoS}

\begin{proof}
By~\cite[Theorem~27.15]{Sat}, the imposed integrability
assumptions,
Condition~(\ref{eq:n0-dep-on-alpha}), and the assumption that $\nu_{0}%
^{\alpha}$ is even combined with a substitution, we have that
\begin{align*}
\int_{|x|>1}e^{\alpha x}\nu(dx)  &  =\int_{|x|>1}e^{\frac{1}{2}\alpha x}%
\nu_{0}^{\alpha}(dx)=\int_{|x|>1}e^{-\frac{1}{2}\alpha
x}\nu_{0}^{\alpha
}(dx)<\infty\,,\\
\int_{|x|>1}e^{x}\nu(dx)  &
=\int_{|x|>1}e^{(1-\frac{1}{2}\alpha)x}\nu
_{0}^{\alpha}(dx)<\infty\,,
\end{align*}
so that, again by~\cite[Theorem~27.15]{Sat}, $\psi_{0}\big(-i(1-\frac{1}%
{2}\alpha)\big)-\psi_{0}\big(i\frac{1}{2}\alpha\big)$ is definable
(and given by the above triplets). It follows that
\begin{align*}
\psi_{0}\big(-i(1-\frac{1}{2}\alpha)\big)-  &  \psi_{0}\big(i\frac{1}{2}%
\alpha\big)=\kappa_{0}\big(1-\frac{1}{2}\alpha\big)-\kappa
_{0}\big(-\frac{1}{2}\alpha\big)\\
&  =(1-\alpha)\frac{1}{2}\sigma^{2}+\int(e^{(1-\frac{1}{2}\alpha)x}%
-e^{-\frac{1}{2}\alpha
x}-x1\hspace*{-0.55ex}\mathrm{I}_{|x|\leq1})\nu
_{0}^{\alpha}(dx)\\
&  =\left(  1-\alpha\right)  \frac{\sigma^{2}}{2}+\int\left(  e^{x}%
-xe^{\frac{1}{2}\alpha
x}1\hspace*{-0.55ex}\mathrm{I}_{|x|\leq1}-1\right)
\,e^{-\frac{1}{2}\alpha x}\nu_{0}^{\alpha}\left(  dx\right)  \,,
\end{align*}
so that by~(\ref{eq:n0-dep-on-alpha}) we arrive
at~(\ref{eq:short-way}).
\end{proof}

\medskip

\section{Exponential L\'{e}vy processes: specific models}

\subsection{Quasi self-dual Normal Inverse Gaussian models}

\label{subs:NIG}

Define $S=e^{\lambda t+X}$, $\lambda\in\mathbb{R}$, with $X$ a
L\'{e}vy process with characteristic function
\begin{equation}
\varphi_{X_{t}}(u)=\exp\Big(t\big(ium+d\big(\sqrt{a^{2}-b^{2}}-\sqrt
{a^{2}-(b+iu)^{2}}\big)\big)\Big)\,, \label{eq:chaf-NIG}%
\end{equation}
with $a>0$, $-a<b<a$, $d>0$, $m\in\mathbb{R}$, as e.g.\
in~\cite[Sections~5.3.8,~5.4]{Sch}. Note that here we do not use
standard Greek letters for the parameters since we intend to
reparameterize the model in order to obtain the quasi self-dual
parameter as a model parameter. The process $X$, introduced by
Barndorff-Nielsen in~\cite{BN,BNa}, and frequently used in the
financial literature, see e.g.~\cite{BN:SHY,con:tan,Sch} and the
literature cited therein, is called Normal Inverse Gaussian (NIG)
process. It can be constructed by time changing a Brownian motion
with drift, is an \textquotedblleft infinite
activity\textquotedblright\ process, and the trajectories of a
NIG-process are of unbounded variation. Furthermore, for the above
parameter restrictions the NIG distribution, i.e.\ the distribution
of $X_{1}$, has semi-heavy tails. The NIG-processes belong to the
class of the Generalized Hyperbolic L\'{e}vy processes. However,
unlike several other processes in this class, the NIG-processes
(along with the Variance Gamma processes, which will be analyzed in
the next section) exhibit the important property that for any
$t\neq1$ the distribution of $X_{t}$ is of the same type as the
distribution of $X_{1}$. This makes the NIG processes preferable
when one works with empirical data, see~\cite{BN:SHY}.

While the centered Gaussian term vanishes, the L\'{e}vy measure of
NIG-processes and the $\gamma$ is given by
\begin{equation}
\nu(dx)=\frac{da}{\pi}\frac{e^{bx}}{|x|}K_{1}(a|x|)dx\,,\quad\gamma
=m+\frac{2da}{\pi}\int_{0}^{1}\sinh(bx)K_{1}(ax)dx\,, \label{eq:nu-NIG}%
\end{equation}
see e.g.~\cite{BNa},~\cite[Sections~5.3.8,~5.4]{Sch}, where $K_{1}$
denotes the modified Bessel function of the third kind with index
$1$, see e.g.~\cite[p.~148]{Sch}.

By further restricting the parameter range for $b$ to the interval
$(-a,a-1)$, cf.\ e.g.~\cite{BN:SHY}, where at the same time we
assume that $a>1/2$ in order to avoid that this interval is empty,
we obtain an existing first exponential moment (for one or
equivalently for all $t>0$, being equivalent to
$\int_{|x|>1}e^{x}\nu(dx)<\infty$, see
e.g.~\cite[Theorem~25.17]{Sat}).

If we rewrite the parameter $b=-\frac{1}{2}\alpha$
then~(\ref{eq:nu-NIG}) reads
\begin{equation}
\nu(dx)=e^{-\frac{1}{2}\alpha
x}\nu_{0}(x)dx\,,\quad\text{with}\quad\nu
_{0}(x)=\frac{da}{\pi|x|}K_{1}(a|x|)\,, \label{eq:nu-NIG-repara}%
\end{equation}
where $\alpha\in(-2(a-1),2a)$. Since $\nu_{0}$ is an even
function, $\nu$ is of the form~(\ref{eq:n0-dep-on-alpha}) so that
it easily follows that $\nu$ satisfies~(\ref{eq:nice-form}).

\begin{proposition}
\label{prop:NIG} Assume that $S=e^{t\lambda+X}$,
$\lambda\in\mathbb{R}$, where $X$ is characterized
by~$(\ref{eq:chaf-NIG})$ with $b=-\frac{1}{2}\alpha$,
$a>\frac{1}{2}$, $\alpha\in(-2(a-1),2a)$, $d>0$. Then

\begin{itemize}
\item[$\mathrm{(i)}$] For $\alpha$ such that
\[
\lambda=-d\big(\sqrt{a^{2}-\frac{1}{4}(2-\alpha)^{2}}-\sqrt{a^{2}-\frac{1}%
{4}\alpha^{2}}\big)\,,
\]
and by subsequently choosing $\gamma$ (via $m=-\lambda$) as
in~$(\ref{Esscher martingale condition-g})$, the asset price model
is quasi self-dual of order $\alpha$ with respect to $\lambda$ and
$e^{X}$ is a martingale.

\item[$\mathrm{(ii)}$] The functions
\[
f_{a,d}:(-2(a-1),2a)\to(-d\sqrt{2a-1},d\sqrt{2a-1})
\]
defined by
\begin{equation}
\label{eq:NIG:eq}f_{a,d}(\alpha)=-d\big(\sqrt{a^{2}-\frac{1}{4}(2-\alpha)^{2}%
}-\sqrt{a^{2}-\frac{1}{4}\alpha^{2}}\big)
\end{equation}
are vanishing if and only if $\alpha=1$, strictly monotonically
decreasing, and bijective with inverse mapping
\[
\alpha_{a,d}:(-d\sqrt{2a-1},d\sqrt{2a-1})\to(-2(a-1),2a)
\]
defined by
\begin{equation}
\label{eq:NIG-solution}\alpha_{a,d}(\lambda) =1-\lambda\frac{\sqrt{4a^{2}%
d^{2}-d^{2}-\lambda^{2}}}{d\sqrt{\lambda^{2}+d^{2}}}\,.
\end{equation}

\end{itemize}
\end{proposition}

Hence, if for given $\lambda$, the parameters $a$, $d$, are chosen
such that $|\lambda|<d\sqrt{2a-1}$ (along with the above
conditions) then we can find the corresponding $\alpha$.
Furthermore, note that~$\mathrm{(ii)}$ implies that $f_{a,d}$ is
non-negative on $(-2(a-1),1]$ and strictly negative on $(1,2a)$,
consistent with Proposition~\ref{prop:jensen}.

\begin{proof}
Under the imposed parameter restrictions we have that $S_{t}$ and
$(S_{t})^{\alpha}$ are integrable for some $t>0$ so that we can
apply Proposition~\ref{prop:short-way-basis}, i.e.\
\begin{multline*}
\lambda=\int\left( e^{x}-xe^{\frac{1}{2}\alpha
x}1\hspace*{-0.55ex}\mathrm{I}_{|x|\leq
1}-1\right)  \,\nu\left(  dx\right) \\
=\psi_{0}\big(-i(1-\frac{1}{2}\alpha)\big)-\psi_{0}\big(i\frac{1}{2}%
\alpha\big)=d(\sqrt{a^{2}-\frac{1}{4}\alpha^{2}}-\sqrt{a^{2}-\frac
{1}{4}(2-\alpha)^{2}})\,,
\end{multline*}
where the last equality is justified by the fact that
$\phi_0(-iv)=e^{\psi_0(-iv)}$, for $v\in(-a,a)$, is the (real
valued) moment generating function which is known in the
literature, see e.g.~\cite{HS}, or which can be derived
from~(\ref{eq:chaf-NIG}) by the standard arguments as in the
proofs of~\cite[Theorems~25.17;~24.11]{Sat} and by setting $b=0$,
$m=0$ and using that $\sinh(0)=0$ so that $\gamma=0$ also holds
(while $\alpha\in(-2(a-1),2a)$ implies that $-\frac{1}{2}\alpha$
and $1-\frac{1}{2}\in(-a,a)$). It remains to show~$\mathrm{(ii)}$.
By $\alpha\in(-2(a-1),2a)$, $a>\frac{1}{2}$, we have
that $a^{2}-\frac{1}{4}(2-\alpha)^{2}>0$ as well as $a^{2}-\frac{1}{4}%
\alpha^{2}>0$, i.e.\ $f_{a,d}$ is differentiable on this interval
(with continuous extension to the closure). Hence,
\[
\lim_{\alpha\rightarrow(-2(a-1))+}f_{a,d}(\alpha)=d\sqrt{2a-1}\,,\quad
\lim_{\alpha\rightarrow(2a)-}f_{a,d}(\alpha)=-d\sqrt{2a-1}\,.
\]
Since $\nu$ satisfies~(\ref{eq:nice-form}), the property that the
function $f_{a,d}$ is strictly monotonically decreasing is a
direct consequence of Proposition~\ref{pr:mon}. Alternatively this
can also be seen by analyzing the derivative. Hence, $f_{a,d}$ is
a bijective mapping from $(-2(a-1),2a)$ to
$(-d\sqrt{2a-1},d\sqrt{2a-1})$. Furthermore, for given $a$, $d$,
$f_{a,d}$ vanishes if and only if $\alpha=1$ (note that due to
$a>\frac{1}{2}$ we always have $1\in(-2(a-1),2a)$).

The calculation for deriving~(\ref{eq:NIG-solution}) is
essentially the same as needed for the results in~\cite{HS}, we
give details in an appendix.

We additionally remark that the equation $m=-\lambda$ for
non-vanishing $\alpha$ is a consequence of the fact that
given~(\ref{Levy quasi self-dual-g}) the condition~(\ref{Esscher
martingale condition-g}) translates into the martingale property
of $e^{\alpha(\lambda t+X_t)}$, so that the corresponding
expectations are identically one. For vanishing $\alpha$ it is a
direct consequence of~(\ref{Esscher martingale condition-g}).
\end{proof}

\subsection{Quasi self-dual Variance Gamma models}

\label{subs:VG}

As already mentioned in the last section, the Variance Gamma (VG)
processes also belong to the class of Generalized Hyperbolic
processes and, as Normal Inverse Gaussian processes, exhibit the
property that for any $t\neq1$ the distribution of $X_{t}$ is of the
same type as the distribution of $X_{1}$. The characteristic
function can be parameterized in different ways, see
e.g.~\cite[Sections~5.3,~5.4]{Sch}. We use here a parametrization
which shows that these processes also belong to the family of
(extended) CGMY processes. Hence, we define $S=e^{\lambda t+X}$,
$\lambda\in\mathbb{R}$, with $X$ being a L\'{e}vy process with
characteristic function
\begin{align}
\varphi_{X_{t}}(u)  &
=\exp\Big(t\big(ium-C(\log(1-\frac{iu}{M})+\log
(1+\frac{iu}{G}))\big)\Big)\nonumber\label{eq:chaf-vg}\\
&
=\exp\Big(t\big(iu\gamma+\int_{\mathbb{R}}(e^{iux}-1-iux1\hspace
*{-0.55ex\mathrm{I}}_{|x|\leq1})\nu(x)\,dx\big)\Big)\,,
\end{align}
where
\begin{equation}
\nu(x)=\frac{C}{|x|}e^{-G|x|}1\hspace*{-0.55ex}\mathrm{I}_{x<0}+\frac{C}%
{|x|}e^{-M|x|}1\hspace*{-0.55ex}\mathrm{I}_{x>0}\,, \label{eq:nu-vg}%
\end{equation}
with $C>0$, $G>0$, $M>0$, along with
\begin{equation}
\gamma=m-\frac{C(G(e^{-M}-1)-M(e^{-G}-1))}{MG}\,. \label{eq:vg-drift-standard}%
\end{equation}
These well-known processes $X$ also appear frequently in the
financial literature, see again e.g.~\cite{BN:SHY,con:tan,Sch} and
the literature cited therein. VG processes possess the
\textquotedblleft infinite-activity\textquotedblright\ property
but have paths of bounded variation.

By choosing
\begin{equation}
M=\beta+\frac{1}{2}\alpha>1\,,\quad\text{ and}\quad G=\beta-\frac{1}{2}%
\alpha>0\,, \label{eq:VG-st-qusi-sd}%
\end{equation}
we ensure that the L\'{e}vy measure is again of the
form~(\ref{eq:nice-form}), more concretely
\[
\nu(x)=e^{-\frac{1}{2}\alpha
x}\nu_{0}(x)\,,\quad\text{with}\quad\nu
_{0}(x)=\frac{Ce^{-\beta|x|}}{|x|}1\hspace*{-0.55ex}\mathrm{I}_{x\neq0}\,,
\]
i.e.\ we have that Condition~(\ref{Levy quasi self-dual-g}) is
satisfied, and in view of Proposition~\ref{pr:mon} that the
integral expression in~(\ref{relation lambda - alpha-g}) is
monotone in $\alpha\in(-2(\beta -1),2\beta)$, where
$\beta>\frac{1}{2}$ is needed in order to ensure that this
interval is not empty. As in Remark~\ref{re:GH-mon} note also that
we have chosen $M>1$, not only $M>0$, in order to ensure that the
first exponential moment along with the characteristic function
(including the L\'{e}vy-Khintchine representation) at the
corresponding point in the complex plane exists,
cf.~e.g.~\cite[Theorem~25.17]{Sat}.

\begin{proposition}
\label{prop:VG} Assume that $S=e^{t\lambda+X}$,
$\lambda\in\mathbb{R}$, where $X$ is characterized
by~$(\ref{eq:chaf-vg})$ with $\alpha\in(-2(\beta -1),2\beta)$,
$\beta>\frac{1}{2}$. Then

\begin{itemize}
\item[$\mathrm{(i)}$] For $\alpha$ such that
\begin{equation}
\label{eq:vg-int-eq-new}\lambda=-C\Big(\log\big(1-\frac{1}{\beta+\frac{1}%
{2}\alpha}\big)+\log\big(1+\frac{1}{\beta-\frac{1}{2}\alpha}\big)\Big)
\end{equation}
and by subsequently choosing $\gamma$ (via $m=-\lambda$) as
in~$(\ref{Esscher martingale condition-g})$ the asset price model
is quasi self-dual of order $\alpha$ with respect to $\lambda$ and
$e^{X}$ is a martingale.

\item[$\mathrm{(ii)}$] The functions
\[
f_{C,\beta}:(-2(\beta-1),2\beta)\rightarrow(-\infty,\infty)
\]
defined by
\begin{equation}
f_{C,\beta}(\alpha)=-C\Big(\log\big(1-\frac{1}{\beta+\frac{1}{2}\alpha
}\big)+\log\big(1+\frac{1}{\beta-\frac{1}{2}\alpha}\big)\Big) \label{eq:VG:eq}%
\end{equation}
are vanishing if and only if $\alpha=1$, are strictly
monotonically decreasing, and bijective with inverse mapping
\[
\alpha_{C,\beta}:(-\infty,\infty)\rightarrow(-2(\beta-1),2\beta)
\]
defined by
\begin{equation}
\alpha_{C,\beta}(\lambda)=%
\begin{cases}
\frac{-2+2\sqrt{e^{-\frac{\lambda}{C}}+\beta^{2}(e^{-\frac{\lambda}{C}}%
-1)^{2}}}{{e^{-\frac{\lambda}{C}}-1}} & \text{for}\ \lambda\neq0\\
1 & \text{for}\ \lambda=0\,.
\end{cases}
\label{eq:VG-solution}%
\end{equation}

\end{itemize}
\end{proposition}

\begin{proof}
Under the imposed parameter restrictions we have that $S_{t}$ and
$(S_{t})^{\alpha}$ are integrable for a $t>0$ so that we can apply
Proposition~\ref{prop:short-way-basis}, i.e.\
\begin{align*}
\lambda &  =\int\left( e^{x}-xe^{\frac{1}{2}\alpha
x}1\hspace*{-0.55ex}\mathrm{I}_{|x|\leq 1}-1\right)  \,\nu\left(
dx\right) =\psi_{0}\big(-i(1-\frac{1}{2}\alpha)\big)-\psi_{0}\big(i\frac{1}{2}%
\alpha\big)\\
&  =-C(\log(1-\frac{1-\frac{1}{2}\alpha}{\beta})+\log(1+\frac{1-\frac{1}%
{2}\alpha}{\beta})-\log(1+\frac{\alpha}{2\beta})-\log(1-\frac{\alpha}{2\beta
}))\,,
\end{align*}
where the last equality can be obtained by a direct calculation or
with the help of the arguments in the Proof of
Proposition~\ref{prop:NIG}, e.g.\ based on~(\ref{eq:chaf-vg}) by
choosing $\alpha=0$ and $m=0$ so that also $\gamma=0$ (and
$M=G=\beta$). Hence,~(\ref{eq:vg-int-eq-new}) follows after
collecting the first and the third as well as the second and the
fourth summand.\newline It remains to show~$\mathrm{(ii)}$.
Obviously, for $\alpha=1$ we have that $f_{C,\beta}$ vanishes. By
$\alpha\in(-2(\beta-1),2\beta)$, we obtain that $f_{C,\beta}$ is
differentiable on this interval. Furthermore, note that
\[
\lim_{\alpha\rightarrow-2(\beta-1)+}f_{C,\beta}(\alpha)=\infty\,,\quad
\lim_{\alpha\rightarrow2\beta-}f_{C,\beta}(\alpha)=-\infty\,.
\]
Since $\nu$ satisfies~(\ref{eq:nice-form}),
Proposition~\ref{pr:mon} implies that $f_{C,\beta}$ is strictly
monotonically decreasing. Alternatively this can also be seen by
the (non-vanishing) derivative on $(-2(\beta-1),2\beta)$ given by
\[
f_{C,\beta}^{\prime}(\alpha)=-\frac{C}{2}\left(  \frac{1}{(\beta+\frac{1}%
{2}\alpha-1)(\beta+\frac{1}{2}\alpha)}+\frac{1}{(\beta-\frac{1}{2}%
\alpha)(\beta-\frac{1}{2}\alpha+1)}\right)  \,,
\]
which is, in view of~(\ref{eq:VG-st-qusi-sd}), $C>0$, obviously
negative on the considered interval.

It remains to show that~(\ref{eq:VG-solution}) is the
corresponding inverse mapping. First note that since the negative
continuous derivative of $f_{C,\beta}$ never vanishes on
$(-2(\beta-1),2\beta)$ we have that the inverse mapping is
continuously differentiable and strictly monotonically decreasing.
Finally, with $C>0$, $M=\beta+\frac{1}{2}\alpha>1$, and
$G=\beta-\frac{1}{2}\alpha>0$,~(\ref{eq:VG:eq}) can be rewritten
as
\[
-\frac{\lambda}{C}=\log\big(1-\frac{1}{\beta+\frac{1}{2}\alpha}\big)
+\log\big(1+\frac{1}{\beta-\frac{1}{2}\alpha}\big)\,,
\]
or equivalently
\[
(e^{-\frac{\lambda}{C}}-1)\alpha^{2}+4\alpha-4(1+\beta^{2}(e^{-\frac{\lambda
}{C}}-1))=0
\]
so that
\[
\alpha=\frac{-2\pm2\sqrt{e^{-\frac{\lambda}{C}}+\beta^{2}(e^{-\frac{\lambda
}{C}}-1)^{2}}} {e^{-\frac{\lambda}{C}}-1}\,,\quad\text{for }
\lambda\neq0\,.
\]
In order to ensure
\[
\lim_{\lambda\to-\infty}\alpha_{C,\beta}(\lambda)=2\beta\,,\quad\lim
_{\lambda\to\infty}\alpha_{C,\beta}(\lambda)=-2(\beta-1)\,,\quad\lim
_{\lambda\to0}\alpha_{C,\beta}(\lambda)=1\,,
\]
we arrive at~(\ref{eq:VG-solution}). Furthermore, note that the
equation $m=-\lambda$ follows by the reasons given in the proof of
Proposition~\ref{prop:NIG}.
\end{proof}

\subsection{Quasi self-dual Meixner models}

\label{sec:Meixner}

In this section we consider share prices modelled by
$S=e^{t\lambda+X}$, $\lambda\in\mathbb{R}$, with $X$, being a
L\'{e}vy process with characteristic function
\[
\varphi_{X_{t}}(u)=e^{i(tm)u}\left(
\frac{\cos(\frac{b}{2})}{\cosh ((au-ib)/2)}\right)  ^{2dt}\,,
\]
where
\[
\nu(dx)=d\frac{e^{\frac{b}{a}x}}{x\sinh{\frac{\pi}{a}x}}\,dx
\]
with $a>0$, $b\in(-\pi,\pi)$, $d>0$, $m\in\mathbb{R}$, see
e.g.~\cite{gri99} or~\cite{Sch} and the literature cited therein.
Again, this process has no centered Gaussian term. The
trajectories of the process $(X_{t})_{t\in \lbrack0,T]}$ have
unbounded variation and the Meixner distribution of $X_{1}$
has semi-heavy tails, see again~\cite{gri99}. In order to ensure $E(e^{X_{1}%
})<\infty$ (and that the characteristic function is
correspondingly extendable) we additionally restrict $b$ to
$(-\pi,\pi-a)$, see e.g.~\cite[Proof~of~Theorem~1]{gri99}, so that
we also have to restrict $a$ to values strictly below $2\pi$ in
order to avoid that this interval is empty. By
keeping the parameters $b$, $d$, and $m$ but writing $\frac{b}{a}=-\frac{1}%
{2}\alpha$ ($b\neq0$ in order to avoid division by zero), the
L\'{e}vy measure reads
\[
\nu(dx)=e^{-\frac{1}{2}\alpha x}\nu_{0}^{\alpha}(x)dx\,,\quad\text{with}%
\quad\nu_{0}^{\alpha}(x)=\frac{d}{x\sinh((-\frac{\alpha\pi}{2b})x)}\,,
\]
with $\nu_{0}^{\alpha}$ being again even in $x$, however, still
depending on
$\alpha$, so that~(\ref{eq:n0-dep-on-alpha}%
) holds, but the conditions in Proposition~\ref{pr:mon} are not
satisfied. Note that the assumption $b\neq0$ immediately implies
that $\alpha$ cannot vanish, an assumption, which is {\it a
priori} not needed. For brevity we focus in the sequel on the
cases $\alpha\neq0$. However, on the basis of
Proposition~\ref{co:quasi PCS-g} it can be derived that the quasi
self-duality of order $\alpha=0$ enforces $b=0$ (so that
$a\in(0,\pi)$ is assumed in order to ensure the existence of the
first exponential moment) and that $m=-\lambda $. Furthermore, the
additional martingale property of $\exp(X)$ for positive carrying
costs can be obtained by ensuring $\lambda=-2d\log(\cos(a/2))$,
for suitably chosen $a\in(0,\pi)$, $d>0$.

Now we discuss the case $\alpha\neq0$ in more detail. In order to
ensure that $a=-\frac{2b}{\alpha}>0$, and that the first
exponential moment exists, we consider the two cases

\begin{itemize}
\item[$\mathrm{(M1)}$] $b\in(0,\pi)$ and $\alpha\in(-\infty,-\frac{2b}{\pi-b}%
)\subset(-\infty,0)$, ($d>0$, $m\in\mathbb{R}$);

\item[$\mathrm{(M2)}$] $b\in(-\pi,0)$ and
$\alpha\in(-\frac{2b}{\pi-b},\infty )\subset(0,\infty)$ ($d>0$,
$m\in\mathbb{R}$).
\end{itemize}

Since $\alpha=0$ is not an admissible parameter for the chosen
parametrization, we take this opportunity to derive here the
risk-neutral self-duality based on
Remark~\ref{re:mart-mart-sym=ok-g}. In our notation and with the
parameter restrictions the characteristic function reads
\begin{equation}
\varphi_{X_{t}}(u)=e^{i(tm)u}\left(
\frac{\cos(\frac{b}{2})}{\cosh
(b\,\frac{2u+i\alpha}{2\alpha})}\right)  ^{2(dt)}\,^{.}
\label{eq:meixner-chaf}%
\end{equation}
In view of the imposed integrability assumptions and the fact that
$X$ is a L\'{e}vy process, it suffices to ensure that
$\varphi_{X_{1}}(-i)=1$, i.e.\
\begin{equation}
m=-2d\log\left(  \frac{\cos(\frac{b}{2})}{\cos(\frac{b}{2}-\frac{b}{\alpha}%
)}\right)  \,, \label{eq:first-m-cond-Meixner}%
\end{equation}
where we refer to the proof of Theorem~1 in~\cite{gri99} for the
fact that the moment generating function at one is of the form of
the r.h.s.\ of~(\ref{eq:meixner-chaf}) for $u=-i$ (for our
parameter restrictions). Now ensure the martingale property of
$S^{\alpha}$. Note that as a consequence
of~\cite[Proposition~11.10]{Sat} or directly with the help
of~(\ref{eq:meixner-chaf}) we obtain that $Y=\alpha\lambda
t+\alpha X$ is again a Meixner process with characteristic
function
\begin{equation}
\varphi_{Y_{t}}(u)=e^{it(\alpha\lambda+\alpha m)u}\left(
\frac{\cos
(\frac{\tilde{b}}{2})}{\cosh(\tilde{b}\,\frac{2u+i}{2})}\right)  ^{2(dt)}%
\end{equation}
with new parameter $\tilde{m}=\alpha(\lambda+m)$, unchanged
parameter $d>0$, and new $\alpha$ now being identically one, as it
should be in the self-dual case. Furthermore, the parameter $b$
turns the sign in the first case~$\mathrm{(M1)}$ and remains
unchanged in the second case~$\mathrm{(M2)}$. Note that in both
cases the first exponential moment of $Y_{t}$ exists. Hence, by
using that $Y$ is a L\'{e}vy process we get the martingale
property of $S^{\alpha}$ by ensuring that $\varphi_{Y_{1}}(-i)=1$,
i.e.\ in both cases
\[
e^{\alpha\lambda+\alpha m}\left(  \frac{\cos(\frac{b}{2})}{\cos(\frac{b}{2}%
)}\right)  ^{2d}=1\,,
\]
i.e.
\begin{equation}
\lambda=-m\,, \label{eq:mart-2-Meixner}%
\end{equation}
where we again refer to the proof of Theorem~1 in~\cite{gri99} as
far as the form of the moment generating function at one is
concerned (note that condition (\ref{eq:mart-2-Meixner}) can also
be derived with the help of the moment generating function of
$X_1$). Hence, in particular in
view of~(\ref{eq:first-m-cond-Meixner}%
,~\ref{eq:mart-2-Meixner}) we obtain the first part of the following
result. In the following we always restrict the $\arccos$ to its
principal branch with range $[0,\pi]$, in particular we then have
for $x\in[-\pi,0]$ that $\arccos(\cos(x))=\arccos(\cos(-x))=-x$.
Furthermore, note that the case~$\mathrm{(M1)}$ and the second
inversion formula in the case~$\mathrm{(M2)}$ do not include the
special case $\alpha=1$.

\begin{proposition}
\label{prop:Meixner} Assume that $S=e^{t\lambda+X}$,
$\lambda\in\mathbb{R}$, where $X$ is characterized
by~$(\ref{eq:meixner-chaf})$ with parameter restrictions given
in~$\mathrm{(M1)}$ or in~$\mathrm{(M2)}$. Then

\begin{itemize}
\item[$\mathrm{(i)}$] For $\alpha$ such that
\[
\lambda=2d\log\left(
\frac{\cos(\frac{b}{2})}{\cos(\frac{b}{2}-\frac
{b}{\alpha})}\right)  \,,
\]
and by subsequently choosing $\lambda=-m$,
Condition~$(\ref{eq:first-m-cond-Meixner})$ is satisfied so that
$e^{X}$ is a martingale and $S$ is quasi self-dual of order
$\alpha$ with respect to $\lambda$.

\item[$\mathrm{(ii)}$] The behavior of the corresponding function
depends on the following cases.

\begin{itemize}
\item[$\mathrm{(M1)}$] The functions
\[
f_{b,d}:(-\infty,-\frac{2b}{\pi-b})\to(0,\infty)
\]
defined by
\begin{equation}
\label{eq:Meixn-functions}f_{b,d}(\alpha)=2d\log\left(  \frac{\cos(\frac{b}%
{2})}{\cos(\frac{b}{2}-\frac{b}{\alpha})}\right)  \,,
\end{equation}
are strictly monotonically increasing, and bijective with inverse
mapping
\[
\alpha_{b,d}:(0,\infty)\to\big(-\infty,-\frac{2b}{\pi-b}\big)
\]
defined by
\begin{equation}
\label{eq:Meixner-inv-M1}\alpha_{b,d}(\lambda)=\frac{2b}{b-2\arccos(\cos
(\frac{b}{2})e^{-\frac{\lambda}{2d}})}\,.
\end{equation}

\item[$\mathrm{(M2)}$] The functions
\[
f_{b,d}:(-\frac{2b}{\pi-b},\infty)\to[2d\log(\cos(b/2)),\infty)
\]
defined by~$(\ref{eq:Meixn-functions})$ are vanishing if and only
if $\alpha=1$, are piecewise injective respectively on
$(-\frac{2b}{\pi-b},2]$ where they are strictly monotonically
decreasing, and on $(2,\infty)$ where they are strictly
increasing. The corresponding inverse mappings are
\[
\alpha_{b,d}:[2d\log(\cos(b/2)),\infty)\to(-\frac{2b}{\pi-b},2]\
\]
again defined by~$(\ref{eq:Meixner-inv-M1})$ as well as
\[
\bar\alpha_{b,d}:(2d\log(\cos(b/2)),0)\to(2,\infty)
\]
here defined by
\begin{equation}
\label{eq:Meixner-exots}\bar\alpha_{b,d}(\lambda)=\frac{2b}{b+2\arccos
(\cos(b/2)e^{-\frac{\lambda}{2d}})}\,.
\end{equation}

\end{itemize}
\end{itemize}
\end{proposition}

\begin{proof}
It remains to show~$\mathrm{(ii)}$. We start with the
case~$\mathrm{(M1)}$. For $\alpha\in (-\infty,\frac{-2b}{\pi-b})$
(note that $0$ is not contained in this interval) we have that
$\frac{b}{2}-\frac{b}{\alpha}\in(\frac{b}{2},\frac{\pi}{2})$ where
the cosine does not vanish. Hence, $f_{b,d}$ is differentiable on
this interval. Furthermore, note that
\[
\lim_{\alpha\rightarrow(-\frac{2b}{\pi-b})-}f_{b,d}(\alpha)=\infty\,,\quad
\lim_{\alpha\rightarrow-\infty}f_{b,d}(\alpha)=0\,,
\]
and that the derivative can be written as
\begin{equation}
f_{b,d}^{\prime}(\alpha)=2d\frac{\sin(\frac{b}{2}-\frac{b}{\alpha})b}%
{\cos(\frac{b}{2}-\frac{b}{\alpha})\alpha^{2}}\,, \label{eq:deriv-Meixner}%
\end{equation}
where for $b\in(0,\pi)$ and since $\frac{b}{2}-\frac{b}{\alpha}\in(\frac{b}%
{2},\frac{\pi}{2})$ all expressions in this quotient are positive
so that the derivative is positive for all
$\alpha\in(-\infty,-\frac{2b}{\pi-b})$, i.e.\ the functions
$f_{b,d}$ are strictly monotonically \emph{increasing}. Now we
derive the corresponding inverse mappings. Since $d>0$ we can
rewrite~(\ref{eq:Meixn-functions}) as
\[
\frac{\lambda}{2d}=\log\left(  \frac{\cos(\frac{b}{2})}{\cos(\frac{b}{2}%
-\frac{b}{\alpha})}\right)
\]
where we note that none of the expressions vanishes for the
imposed parameter restrictions. This is equivalent to
\[
\cos(\frac{b}{2}-\frac{b}{\alpha})=\cos(\frac{b}{2})e^{-\frac{\lambda}{2d}}%
\]
where we note that $e^{-\frac{\lambda}{2d}}<1$ for the present
parameter restrictions and $\lambda>0$. Since
$\frac{b}{2}-\frac{b}{\alpha}\in(\frac {b}{2},\frac{\pi}{2})$ this
yields
\begin{equation}
\frac{b}{2}-\frac{b}{\alpha}=\arccos(\cos(\frac{b}{2})e^{-\frac{\lambda}{2d}%
})\,. \label{eq:difference-Meixner}%
\end{equation}
By the monotonicity property of $\arccos$ we have that
$\arccos(\cos(\frac
{b}{2})e^{-\frac{\lambda}{2d}})\neq\frac{b}{2}$ for the present
parameter restrictions and $\lambda>0$ so that we end up
with~(\ref{eq:Meixner-inv-M1}) and we finally note that
\[
\lim_{\lambda\rightarrow0+}f_{b,d}(\lambda)=-\infty\quad\text{ and
}\quad
\lim_{\lambda\rightarrow\infty}f_{b,d}(\lambda)=-\frac{2b}{\pi-b}\,.
\]
\newline For the cases~$\mathrm{(M2)}$ we first observe that for the imposed parameter
restrictions the functions $f_{b,d}$ are differentiable on
$(-\frac{2b}{\pi -b},\infty)$ with derivative given
in~(\ref{eq:deriv-Meixner}). However, here we have that for $b<0$
with $\alpha\in(-\frac{2b}{\pi-b},2)$, $\sin(\frac
{b}{2}-\frac{b}{\alpha})$ along with the other remaining
expressions is still positive so that the derivatives are negative
and the functions $f_{b,d}$ monotonically decreasing. However,
where for $\alpha\in(2,\infty)$ the sine turns its sign so that
the derivatives become positive, i.e.\ the functions are
monotonically increasing there, the derivative vanishes at the
minimum of $f_{b,d}$ in $\alpha=2$. As far as the inversions are
concerned, we remark
that for $\lambda\geq2d\log(\cos(\frac{b}{2}))$ we get $\cos(\frac{b}%
{2})e^{-\frac{\lambda}{2d}}\leq1$. Since for
$\alpha\in(-\frac{2b}{\pi-b},2]$ we have that
$\frac{b}{2}-\frac{b}{\alpha}$ is non-negative, the inversion
follows by the same steps as in the case~$\mathrm{(M1)}$ so that~(\ref{eq:Meixner-inv-M1}%
) again follows. However, for $\alpha>2$ we have that
$\frac{b}{2}-\frac {b}{\alpha}<0$ so that we obtain
\[
\frac{b}{\alpha}-\frac{b}{2}=\arccos(\cos(\frac{b}{2})e^{-\frac{\lambda}{2d}%
})
\]
instead of~(\ref{eq:difference-Meixner}) so that we finally end up
with~(\ref{eq:Meixner-exots}) (where we stress that
$\lambda\in(2d\log (\cos(b/2)),0)$). To conclude, we remark that
\[
\lim_{\lambda\rightarrow\infty}\alpha_{b,d}(\lambda)=-\frac{2b}{\pi
-b}\,,\
\lim_{\lambda\rightarrow2d\log(\cos(b/2))+}\bar{\alpha}(\lambda
)=2\,,\ \lim_{\lambda\rightarrow0-}\bar{\alpha}(\lambda)=\infty\,,
\]
while $\alpha_{b,d}(2d\log(\cos(b/2)))=2$.
\end{proof}


\section*{Appendix: Inverse in the Normal Inverse Gaussian case}
First note that due to $|\lambda|<d\sqrt{2a-1}$ (and $d>0$) as
well as because of $a>\frac{1}{2}$ we have
\[
\lambda^{2}<d^{2}(2a-1)<4a^{2}d^{2}-d^{2}\,,\quad\text{equivalently}
\quad4a^{2}d^{2}-d^{2}-\lambda^{2}>0\,.
\]
Furthermore, $d>0$ implies that $d\sqrt{\lambda^{2}+d^{2}}$ is
non-vanishing and that
\[
g_{a,d}(\lambda)=\frac{\sqrt{4a^{2}d^{2}-d^{2}-\lambda^{2}}}{d\sqrt
{\lambda^{2}+d^{2}}}>0 \quad\text{for all}\quad\lambda\in(-d\sqrt{2a-1}%
,d\sqrt{2a-1})\,,
\]
i.e.\
\begin{equation}
\label{eq:help-for-square}\alpha_{a,d}(\lambda)=1-\lambda
g_{a,d}(\lambda)\,,
\ \text{ with }\ g_{a,d}>0\ \text{ for all }\lambda\in(-d\sqrt{2a-1}%
,d\sqrt{2a-1})\,.
\end{equation}
The function $\alpha_{a,d}$ is differentiable on
$(-d\sqrt{2a-1},d\sqrt {2a-1})$ (this could also be seen by
noticing that the derivative of $f_{a,d}$ does not vanish and by
the inversion below), with derivative
\[
\alpha_{a,d}^{\prime}(\lambda)=-\frac{4a^{2}d^{4}-(d^{2}+\lambda^{2})^{2}}
{d(\lambda^{2}+d^{2})^{(3/2)}\sqrt{4a^{2}d^{2}-d^{2}-\lambda^{2}}}\,,
\]
where, again by $|\lambda|<d\sqrt{2a-1}$, $a>\frac{1}{2}$, we
obtain $(\lambda^{2}+d^{2})^{2}<(2ad)^{2}=4a^{2}d^{4}$, i.e.\
$\alpha_{a,d}$ is strictly monotonically decreasing on
$(-d\sqrt{2a-1},d\sqrt{2a-1})$ with
\[
\lim_{\lambda\to(-d\sqrt{2a-1})+}\alpha_{a,d}(\lambda)=2a\,,\quad\lim
_{\lambda\to(d\sqrt{2a-1})-}\alpha_{a,d}(\lambda)=-2(a-1)\,.
\]
Furthermore, by $d>0$,~(\ref{eq:NIG:eq}) can be rewritten as
\[
\sqrt{a^{2}-\frac{1}{4}\alpha^{2}}-\frac{\lambda}{d}=\sqrt{a^{2}-\frac{1}%
{4}(2-\alpha)^{2}}\,,
\]
where we recall that $a^{2}-\frac{1}{4}\alpha^{2}>0$, and
$a^{2}-(1-\frac {1}{2}\alpha)^{2}>0$ for $\alpha\in(-2(a-1),a)$,
by squaring and rearranging, this implies
\[
(1+(\frac{\lambda}{d})^{2})-\alpha=2\frac{\lambda}{d}\sqrt{a^{2}-\frac{1}%
{4}\alpha^{2}}\,.
\]
Squaring again, multiplying by $d^{4}$, and rearranging yields
\[
(d^{2}(d^{2}+\lambda^{2}))\alpha^{2}-2d^{2}(d^{2}+\lambda^{2})\alpha
+((d^{2}+\lambda^{2})^{2}-4a^{2}\lambda^{2}d^{2})=0\,.
\]
Again by noticing that $d>0$, $d^{2}(d^{2}+\lambda^{2})>0$, and $4a^{2}%
d^{2}-d^{2}-\lambda^{2}>0 $ we see that the solution needs to be
of the form
\begin{align*}
\alpha_{1,2}(\lambda)  &  =
\frac{2(d^{2}+\lambda^{2})d^{2}\pm\sqrt
{4d^{4}(d^{2}+\lambda^{2})^{2}-4d^{2}(d^{2}+\lambda^{2}) ((\lambda^{2}%
+d^{2})^{2}-4a^{2}\lambda^{2}d^{2})}}{2d^{2}(d^{2}+\lambda^{2})}\\
&  =1\pm\frac{\sqrt{4d^{2}(d^{2}+\lambda^{2})\lambda^{2}(4a^{2}d^{2}%
-\lambda^{2}-d^{2})}}{2d^{2}(\lambda^{2}+d^{2})}\\
& =1\pm|\lambda|\frac{\sqrt{4a^{2}d^{2}-d^{2}-\lambda^{2}}}{d\sqrt
{d^{2}+\lambda^{2}}}\,.
\end{align*}
Since $g_{a,d}(\lambda)>0$ for all $|\lambda|<2\sqrt{2a-1}$ (where
$a>\frac {1}{2}$) and because the function is decreasing in
$\lambda$ we end up with~(\ref{eq:NIG-solution}).

\end{document}